\documentclass[prd,twocolumn,amsmath,amssymb,floatfix,superscriptaddress]{revtex4-2}

\include{pack}

\newcommand{\physrep}{{Physics~Reports}}

\newcommand{\aap}{{Astron.~Astrophys.}}

\def\aap{A\&A}

\def\physrep{Phys. Rep.}

\graphicspath{{./fig/}}

\begin{document}

\title{Non-parametric modeling of cosmological data base on the $\chi^2$ distribution}

\author{Maryam Vazirnia }
\affiliation{Department of Physics, Bu-Ali Sina University, Hamedan
	65178, 016016, Iran}

\author{Ahmad Mehrabi }
\affiliation{Department of Physics, Bu-Ali Sina University, Hamedan
65178, 016016, Iran}

\begin{abstract}
In the $\Lambda$CDM model, cosmological observations from the late and recent universe reveal a puzzling $\sim 4.5\sigma$ tension in the current rate of universe expansion. In addition to the various scenarios suggested to resolve the tension, a non-parametric modeling may provide useful insights. In this paper, we look at three well-known non-parametric methods, the smoothing method, the genetic algorithm and the Gaussian process. Considering these three methods, we employ the recent Hubble parameters data to reconstruct the rate of universe expansion and supernovae Pantheon sample to reconstruct the luminosity distance. In contrast to the similar studies in the literature, the chi-squared distribution has been used to construct a reliable criterion to select  a  reconstruction. Finally, we compute the current rate of universe expansion ($H_0$) for each method, provide some discussions regarding performance of each approach and compare the results.
\end{abstract}
\maketitle

\section{Introduction}

 The analysis of supernova type Ia (SNIa) data in \cite{Riess_1998,Perlmutter_1999} revealed for the first time that the expansion of the universe is accelerating. Several independent measurements since then, including the cosmic microwave background (CMB) \cite{Planck:2015xua,Aghanim:2018eyx}, baryon acoustic oscillation (BAO)\cite{Reid:2012sw,Abbott:2017wcz,Alam:2016hwk,Gil-Marin:2018cgo}, cosmic chronometers \cite{Stern_2010,Farooq:2016zwm} and large scale structures \cite{Alam_2017,Abbott_2018,ebosscollaboration2020completed} confirmed such an exotic behavior of the universe at the recent time. Many attempts have been made to describe this phenomena after these confirmations. There are mainly two avenues to find a solution, an exotic matter with negative equation of state (EoS) \cite{Copeland:2006wr,Chiba:2009nh,Amendola:2010,Mehrabi:2018dru,Mehrabi:2018oke} or modification of gravity at large scales\cite{Magnano:1993bd,Dobado:1994qp,Capozziello:2003tk,Carroll:2003wy,Sebastiani:2016ras}. Among all the alternatives, a cosmological constant with EoS $w=-1$ and cold dark matter, the so called $\Lambda$CDM, gives a good model to describe practically all data sets. However, the model suffers from severe theoretical problems which have not been resolved yet \cite{2003PhR...380..235P,perivolaropoulos2008puzzles,padilla2015lectures}. 

Considering a Friedman universe, the present rate of expansion ($H_0$) is one of the most essential characteristics in a cosmological scenario. Unfortunately, there is a strange discrepancy between measurement of $H_0$ considering late and early cosmological data. Assuming the $\Lambda$CDM model, the CMB data yields a relatively lower value of $H_0\sim 67 Kms^{-1}Mps^{-1}$ \cite{Aghanim:2018eyx}, whereas local SNs anticipate $H_0\sim 74 Kms^{-1}Mps^{-1}$ \cite{Riess:2019cxk}. It's worth noting that the former requires a model to measure $H 0$, whereas the latter relies solely on a distance ladder and the absolute magnitude of SNs. There have been numerous attempts to tackle this problem ( see\cite{divalentino2021realm} and its references) but no adequate solution has yet been found.

Given a data set, a model with some free parameters can be chosen to describe the data. The free parameters should be constrained using a statistical method in this circumstance. In terms of statistics, the frequentiest and Bayesian scenarios are the two options for constraining the free parameters.  In the frequntiest point of view, some estimators have been defined to estimate the values of parameters and their uncertainty. In the Bayesian scenario, on the other hand, the Bayes theorem has been used to determine the posterior probability distribution of the parameters. The distribution was then utilized to determine the optimal parameter values as well as their uncertainty \cite{Trotta:2017wnx}. 

On the other hand, using a non-parametric (NP) approach to describe a data collection is a viable option. In machine learning scenarios, this type of modeling was created to have as much capacity to describe a data set as possible. A NP is a unique way to investigate a data collection without having to assume any particular parametric shape. Furthermore, the NP generates a large number of reconstructions at once, some of which may have novel features. The Gaussian process (GP) is the most well-known NP modeling in cosmology, and it has been frequently employed in cosmic data analysis \cite{Shafieloo_2012,Liao_2019,G_mez_Valent_2018,Pinho_2018,Mehrabi_2020,Dhawan:2021mel,Escamilla-Rivera:2021rbe,Bernardo:2021mfs,Briffa:2020qli}. In this case, the data has been modeled by a sequence of  Gaussian random values. The Genetic algorithm (GA) on the other hand, given a set of base functions, can reconstruct multiple curves that are consistent with the data. This method has been used in \cite{Bogdanos_2009} to study SNIa data and in \cite{Nesseris_2010} to investigate a null test on the cosmological constant. Moreover, this method has been used to study the dark energy in \cite{Nesseris_2012}. 
The smoothing method (SM) is the last procedure, which involves reconstructing a function using a smoothing kernel. In this case, a series of initial guesses at random sites has been improved with each iteration, resulting in a better fit to the data after each step.  This method was utilized in \cite{Shafieloo_2007,Shafieloo_2010,Shafieloo_2018} to investigate various cosmological data. 

We select these methods because they all rely on a sample of reconstructions, allowing us to compare them. However, with an NP approach, estimating errors is a challenging issue, which is a drawback in such cases. Reconstructions in the GP are obtained by sampling from a multivariate Gaussian distribution, with the standard deviation (mean) of a quantity at each point indicating the quantity's uncertainty (central value). In contrast to the GP, there is no unique way to select a reconstruction and make a sample in the GA and SM. Considering SM method, in \cite{Shafieloo_2007,Shafieloo_2010,Shafieloo_2018} a reconstruction with $\chi^2<\chi^2_{\rm ref}$ has been selected where $\chi^2_{\rm ref}$ is the $\chi^2$ of a reference model e.g the $\Lambda$CDM. On the other hand, considering the GA, in \cite{Nesseris_2010,Nesseris_2012,Arjona:2020axn,Arjona:2020skf}, only the best reconstruction was obtained, and the uncertainty of that reconstruction was calculated in a different approach.  In this paper, we introduce a more reliable sampling method to estimate the central value, with optimized chi-squared values that select the reconstructions.
 
The structure of this paper is as follows: We present the statistical tools needed for our analysis in section (\ref{sect:parameter_inference}), as well as the chi-squared distribution and our new criterion for selecting a reconstruction in NP modeling. A brief discussion of parameter inference in a model dependent technique has also been included. All three NP approaches are described in depth in section (\ref{sect:non-par}). Furthermore, given a data set, we discuss how each method provides a consistent reconstruction. In section (\ref{sect:res}), we present the results for different methods and discuss how different criterion affect the estimation of $H_0$. Finally, in section (\ref{conclude}), we provide our findings, explain and compare them to similar efforts in the literature.

\section{Parameter inference and $\chi^2$ distribution}\label{sect:parameter_inference}
Given a data set as $(x_i,y_i,\sigma_i)$ and knowing the distribution of noise, it is an easy task to construct a likelihood function. The likelihood contains all information relating to observed data and is a necessary requirement to infer the free parameters of a model. The Gaussian likelihood is given by,
\begin{equation}\label{eq:likelihood}
 \mathcal{L}(\vec{\theta}) \propto \Pi_{i=1}^{N}\exp({-\frac{1}{2}\frac{(f(x_i,\vec{\theta})-y_i)^2}{\sigma_i^2}}),
\end{equation}      
where $f(x_i,\vec{\theta})$ is the model prediction at $x_i$, $\vec{\theta}$ indicates all free parameters and $N$ is the number of data points. It is common to write the likelihood as:
\begin{equation}\label{eq:log-likelihood}
\mathcal{L}(\vec{\theta}) = \mathcal{L}_0\exp{(-\frac{1}{2}\chi^2)},
\end{equation}
where $\chi^2 = \sum_{i=1}^{N}\frac{(f(x_i,\vec{\theta})-y_i)^2}{\sigma_i^2}$ and $\mathcal{L}_0$ is a normalization constant.  Since the likelihood is Gaussian, the quantity $X_i=\frac{f(x_i,\vec{\theta}-y_i)}{\sigma_i}$ is a Gaussian random variable with zero mean and $var=1$. Summing up the square of $X_i$ gives a quantity $Q=\sum_{i=1}^{k}X_i^2$, which is distributed according to the chi-squared probability distribution function (PDF) for $x=\chi^2$,
\begin{equation}
P(x|k)=\frac{x^{\frac{k}{2}-1}e^{-\frac{x}{2}}}{2^{\frac{k}{2}\Gamma(\frac{k}{2})}}.
\end{equation}
Where $k$ is the number of the degree of freedom (NDF). Furthermore, the PDF shows a peak at $\chi^2\sim k$ for a moderately big $k$ and the probability for $\chi^2>>k$ and $\chi^2<<k$ are negligible. In fact, if $\chi^2>>k$, majority of data points are away from the model prediction, indicating under-fitting case. On the other hand,  if $\chi^2<<k$, majority of the data points are unexpectedly near to the model prediction, we have a over-fitting case. It's worth noting that neither of these scenarios is a good fit for the data. The chi-squared PDF could be used as a criterion to select a reconstruction based on its probability, according to the aforementioned reasoning. Given a percentage number, such as $95\%$, it is straightforward to compute the range ($\chi^2_{min},\chi^2_{max}$) of the interval that lie within this percentage.
\begin{equation}\label{eq:chi2-pro_max}
Pr(\chi^2_{min}<x<\chi^2_{max})=\int_{\chi^2_{min}}^{\chi^2_{max}}dxP(x|k).
\end{equation}
The probability of a reconstruction with a $\chi^2$ within the interval $(\chi^2_{min},\chi^2_{max})$ is equal to the $Pr$. Reconstructions in a sample chosen based on the chi-squared are free from both under-fitting and over-fitting, indicating that the sample is reliable for estimating the value of a parameter. As we mentioned above, authors of \cite{Shafieloo_2012_2,10.1093/mnras/sty398,Shafieloo_2018}, utilized the condition $\chi^2<\chi^2_{\rm ref}$ to select a reconstruction for the SM technique. In such a procedure, there might be some over-fitting reconstructions, making the sample not reliable for estimation of a quantity. 

In contrast to a NP modeling, for a parametric modeling, observational data can be used to constrain its parameters.     
In frequentist scenario, one might use the maximum likelihood estimator (MLE) to estimate the optimal values of the parameters. There are several ways for determining the MLE given a likelihood function. In addition, expanding the likelihood around its peak gives an estimation of the parameter uncertainties \cite{Trotta:2017wnx}. In contrast to this point of view, the Bayes theorem may be used to find the PDF of parameters and from the PDF, the best values and their uncertainties can be easily determined. The Bayes theorem is given by:
\begin{equation}
P(\vec{\theta}|d)= \frac{\mathcal{L}(\vec{\theta}) P(\vec{\theta})} {P(d)},
\end{equation}\label{eq:bays}
where $P(\vec{\theta})$ is the prior information on parameters and $P(d)$ is the Bayesian evidence. With the exception of a few cases that have an analytic solution, the majority of problems should be addressed numerically. In these cases, a numerical approach such as Markov Chain Monte Carlo (MCMC) has been used to estimate the posterior $P(\vec{\theta}|d)$ by sampling from the numerator of the above equation.

\section{NP modeling of a data set}\label{sect:non-par}
 In machine learning domain, a NP approach is used to have as much as possible capacity to describe a data set. The NP methods have been used in number of studies (some of them are listed in introduction) in cosmology to describe a data set. Because little is known about dark energy, using NP modeling could provide new insights or show a new feature in the data. In this work, we consider three NP methods, all reliant upon a sample of reconstruction and compare the results. Since the results depends on the sample, a reliable selection criterion is required.
In this section, we'll go over the basics of three well-known NP methods: GP, GA, and SM. Then, in the following section, we compare the results after finding all viable reconstructions using two separate cosmological data sets. This study helps us understand the performance of each method as well as advantage and disadvantage of each approach. 

\subsection{Gaussian process}   

A Gaussian process is a sequence of Gaussian random variables (RV) that can be presented by a multivariate Gaussian distribution. In this case, the diagonal entries represent uncertainty at each point, whereas off-diagonal terms represent correlation between points. Assume that a data set can be modeled by a Gaussian process, we have
\begin{equation}
f(x)\sim GP(\mu(x),K(x,\tilde{x})),
\end{equation}  
 where $K(x,\tilde{x})$ is the kernel function and $x$, $\tilde{x}$ are two different observational points. The $\mu(x)$ is the mean function which provides the mean of Gaussian RV at each observational point. The kernel gives the covariance matrix of the multivariate Gaussian distribution and depends on some hyper-parameters. The most well known kernel is the squared exponential given by,
$$K(x,\tilde{x}) = \sigma_f^2\exp{\frac{-(x-\tilde{x})^2}{2l^2}}$$ where $\sigma_f^2$ and $l$ are two hyper-parameters.
Given a GP at some observational points $x$, it is simple to find the GP at some arbitrary points $x^{\star}$. The function's values at these points are given by sampling from a multivariate Gaussian $N(\mu^{\star},\Sigma^{\star})$, where the mean and covariance are given by \cite{10.5555/1162254}:

\begin{eqnarray}\label{eq:GP}
\mu^{\star} &=& K(x,x^{\star})[K(x,x^{\star})+C_D]^{-1}Y\\
\Sigma^{\star} &=& K(x^{\star},x^{\star}) - K(x^{\star},x)[K(x,x^{\star})+C_D]^{-1}K(x,x^{\star}) \nonumber
\end{eqnarray}

where $C_D$ is the covariance matrix of the data and $Y=[y_i]$ is the column vector of observational data (as the previous section, we assume a data set as $(x_i,y_i,\sigma_i)$). Notice that, in the above equations, a zero mean prior has been considered for the $\mu^{\star}$.

 Given a data set, one can easily compute the mean and covariance of the multivariate Gaussian distribution and then by sampling from that find many reconstructions. Notice that since the derivative of a GP is another GP, it is also easy to find the derivative of reconstructed function.(to see more details refer to \cite{10.5555/1162254,Mehrabi_2020}). We use the {\it 'GaussianProcessRegressor'} class of the {\it{scikit learn}} to find the mean and covariance matrix of a GP \cite{scikit-learn}. 

 \subsection{Genetic algorithm}
 
 GA is an optimization method inspired by the process of natural selection. it relies on the biologically inspired operators such as crossover, mutation and selection. The process starts from a population of individuals and produces a new generation through mutation, which is a random change in an individual, and crossover, which is a combination of many individuals. In this scenario, the probability of next generation is given by a fitness function which is also our objective function for optimization. For more details on the algorithm, we refer the reader to \cite{Bogdanos_2009}. 

One of the methods of GA is the symbolic regression (SR). In this case, the method attempts to produce a mathematical expressions to describe a data set. To do so, the SR generate an initial random population of individuals using a set of basic functions. Then individuals evolve through GA process to find a new generation with a smaller fitness function. The process repeats until either a minimum fitness function or a certain number of generations is reached. We set the fitness function to the $\chi^2$ and hence the next generation provides a mathematical expression with smaller $\chi^2$. In our analysis, we use the public package {\it{gplearn}} which is an extension of {\it{scikit-learn}} machine learning library to perform the SR. We have examined the hyper-parameters space empirically to find the optimum values. In our code, we have set the tournament size=30, probability of mutation=0.03, probability of crossover=0.9 and population-size = 2000. In addition, we only use "add", "subtract" and "multiplication" for the base functions. 

 The GA method has been used to study the expansion of universe in \cite{Arjona:2019fwb}, dark energy anisotropic stress in  \cite{Arjona_2020}, the cosmic distance duality relation in \cite{arjona2020machine} and null tests for the spatial curvature and homogeneity of the Universe in \cite{arjona2021novel}. However, in these works authors found the best fit among all reconstructions and estimated its uncertainty using some methods base on the path integral. In contrary, in our analysis, we run our code with several different random seeds to generate many reconstructions and then use our new selection criterion to make a reliable sample.

 \subsection{SM method}
In addition to the above approaches, SM is also a tool to study a data set in a model independent manner. It has only a hyper-parameter, the smoothing width $\Delta$, which depends on the number of data points as well as their qualities. In this scenario, SM begins with a sequence of arbitrary guess values and try to generate a new smooth curve that is closer to the data points at each step.  

Given the $C_D$, the covariance of data, to reconstruct a function at arbitrary $x^{\star}$ points, one can start from an initial guess and find the improved values $(\hat{f}_{n+1}(x^{\star}))$  at the next iteration by:
\begin{equation}\label{eq:scal-fac}
	\hat{f}_{n+1}(x^{\star}) = \hat{f}_{n}(x^{\star})+{{\delta f_n^T .C_{D}^{-1}.W(x^{\star})} \over {1^T.C_{D}^{-1}.W(x^{\star})}}, 
\end{equation} 
where the kernel function $W(x^{\star})$ and $\delta f_n^T$ are given by:
\begin{equation}\label{eq:scal-fac2}
	W_i(x^{\star}) = exp{   ({{- \ln^2{{(1+x^{\star}) \over{(1+x_i)}} } \over {2 \Delta^2}})} }, 
\end{equation} 
\begin{equation}\label{eq:scal-fac3}
	\delta f_n = f(x_i) - \hat{f}_{n}(x_i), 
\end{equation} 
the $1^T$ is a unite column vector (its size is the same as the size of data) and $f(x_i)=y_i$ are the observed values. Notice that similar to the GP, $x_i$ ($x^{\star}$) indicates observational (arbitrary) points. This method was used in \cite{Shafieloo_2012_2,10.1093/mnras/sty398,Shafieloo_2018} to reconstruct  the expansion history using cosmological data. However, in these works, authors chose reconstructions under condition $\chi^2<\chi^2_{\Lambda CDM}$ to make a sample. Since the method relies on a sample of reconstructions, we also consider this scenario to investigate our data sets. In contrast to the previous works, we use a different strategy to set the initial guess. The Taylor expansion of the Hubble parameter up to the fourth order in $z$ gives us a Hubble parameter in terms of $(H_0,q_0,j_0,l_0)$ (the cosmography parameters), which is independent of any model \cite{Feeney:2018mkj,Risaliti:2018reu,Lusso_2020,Rezaei_2020} (To see the expansion coefficients as a function of cosmography parameters refer to \cite{Rezaei_2020}). Notice that in order to expand the range of the convergence,  the redshift has changed to the so called y-redshift $y=\frac{z}{1+z}$.
We sample $(H_0,q_0,j_0,l_0)$ from a wide uniform distribution for each initial guess, then utilize them to generate the initial Hubble parameter at $x^{\star}$. Note that, we have double-checked the results to ensure that they are robust and unaffected by the initial guess.           

\section{Data set and results}\label{sect:res}
In order to investigate output of above mentioned NP methods, we consider two separate cosmological data. At the background level, the luminosity and angular diameter distances, as well as direct measurements of the Hubble parameter, have provided information about the universe's expansion rate. From the luminosity and angular diameter distances, one can measure the source comoving distance and then derive the Hubble parameter. Assuming a flat geometry, the Hubble parameter is given by,
\begin{eqnarray}\label{eq:hub-com}
D(z) &=& \int_0^z \frac{dz}{H(z)} \nonumber \\
H(z) &=& \frac{1}{D'(z)},  
\end{eqnarray}
where $D(z)$ is the comoving distance at redshift $z$. Notice that, if $D'(z)=0$ at a point, we can't use the equation to find the Hubble parameter. For all reconstructions of the comoving distance, we check this condition and discard those that have $D'(z)=0$ at one or more points.  

\begin{figure}[h]
	\centering
	\includegraphics[width=8.2 cm]{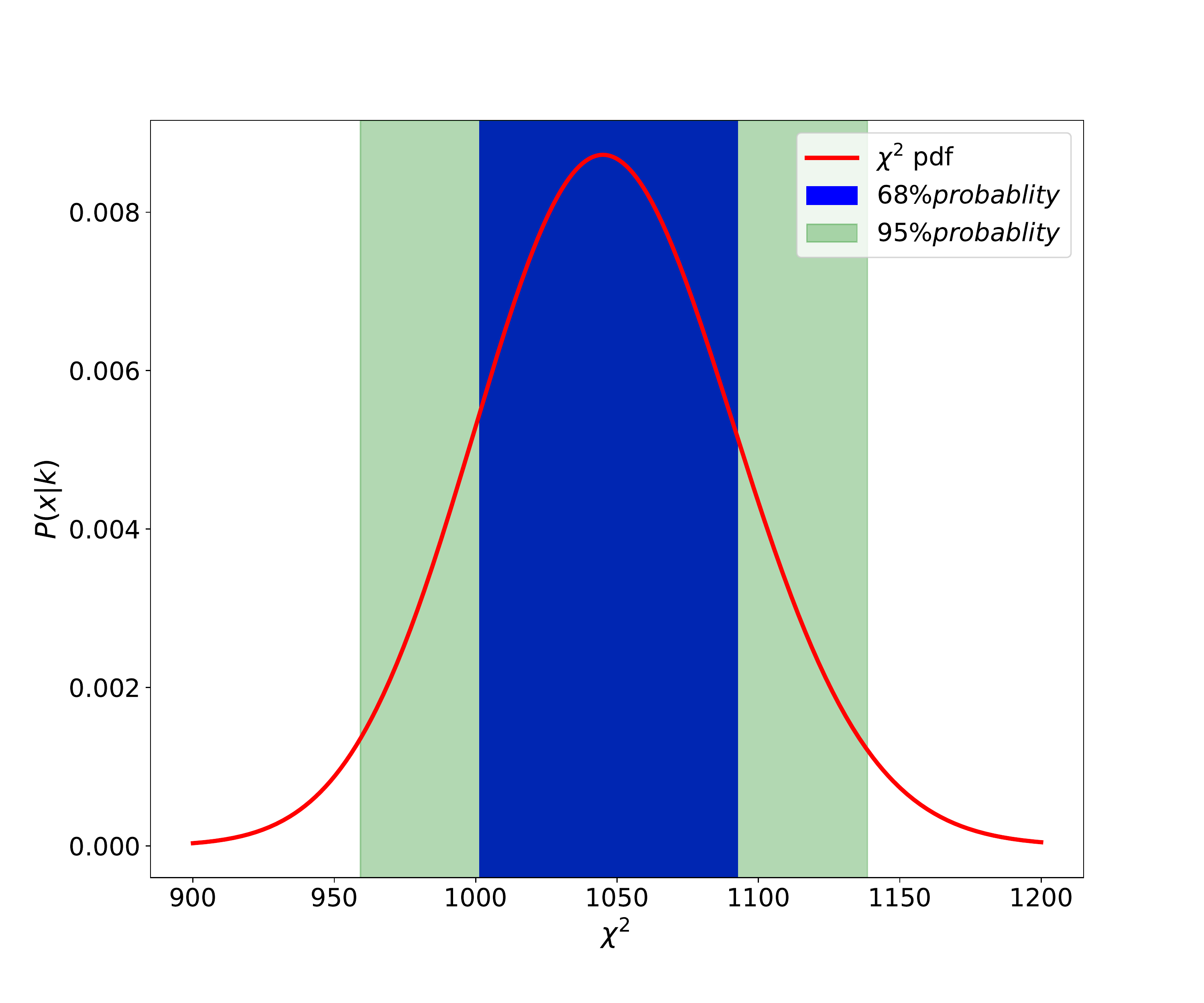}
	\caption{The chi-squared PDF (solid red line) for SNIa data and the range of $\chi^2$ for $pr=68\%$ and $pr=95\%$. }
	\label{fig:chi2_sn}
\end{figure}

The most up-to-data and precise measurement of luminosity distance at this moment is the SNIa Pantheon sample \cite{Scolnic_2018}. This sample contains 1048 spectroscopically confirmed SNIa up to redshift $z=2.26$. To avoid the degeneracy between $H_0$ and the absolute magnitude of the SNIa, we set $M=-19.3$ through our study. Using this data set, we generate a large number of reconstructions through each method, then we used the chi-squared PDF with probability $pr=68\%$ $pr=95\%$ to make a sample. These two intervals for the SNIa data are presented in Fig.(\ref{fig:chi2_sn}). Finally using Eq.(\ref{eq:hub-com}), these reconstructions converted to the Hubble parameter. 

 The second data set is the measurement of the Hubble parameter data collected in \cite{Farooq:2016zwm}. The data set includes measurement of the cosmic chronometer as well as the radial BAO. In addition to the 38 data points in the collection, we add the measurement of $H_0$ from nearby SNs \cite{Riess:2019cxk}. Note that the BAO data points are correlated but for the sake of simplicity, we ignore such correlations in current study.

 \subsection{Results for the SNIa data} 
 
 \begin{figure*}[h]
 	\centering
 	\includegraphics[width=8.2 cm]{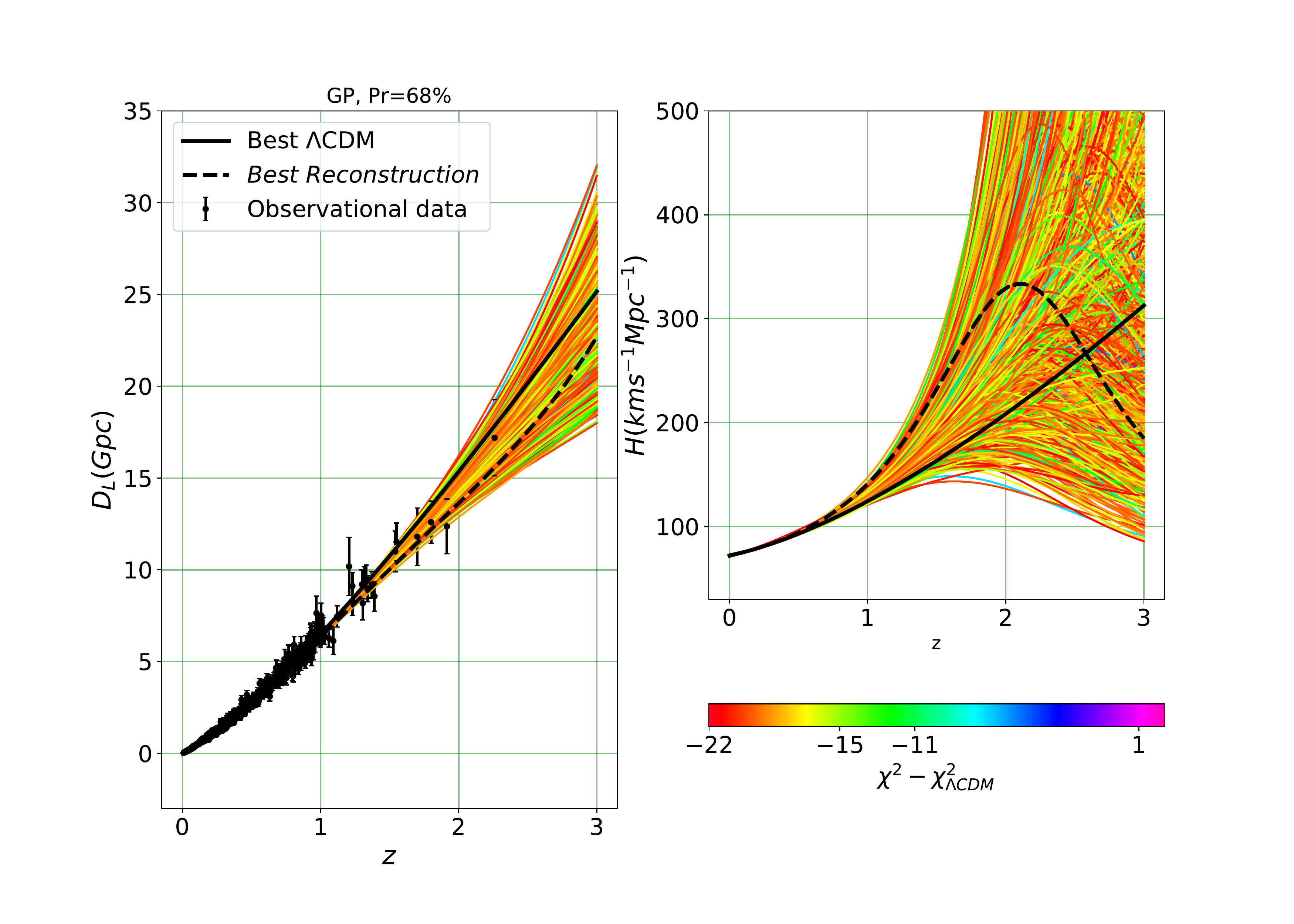}	\includegraphics[width=8.2 cm]{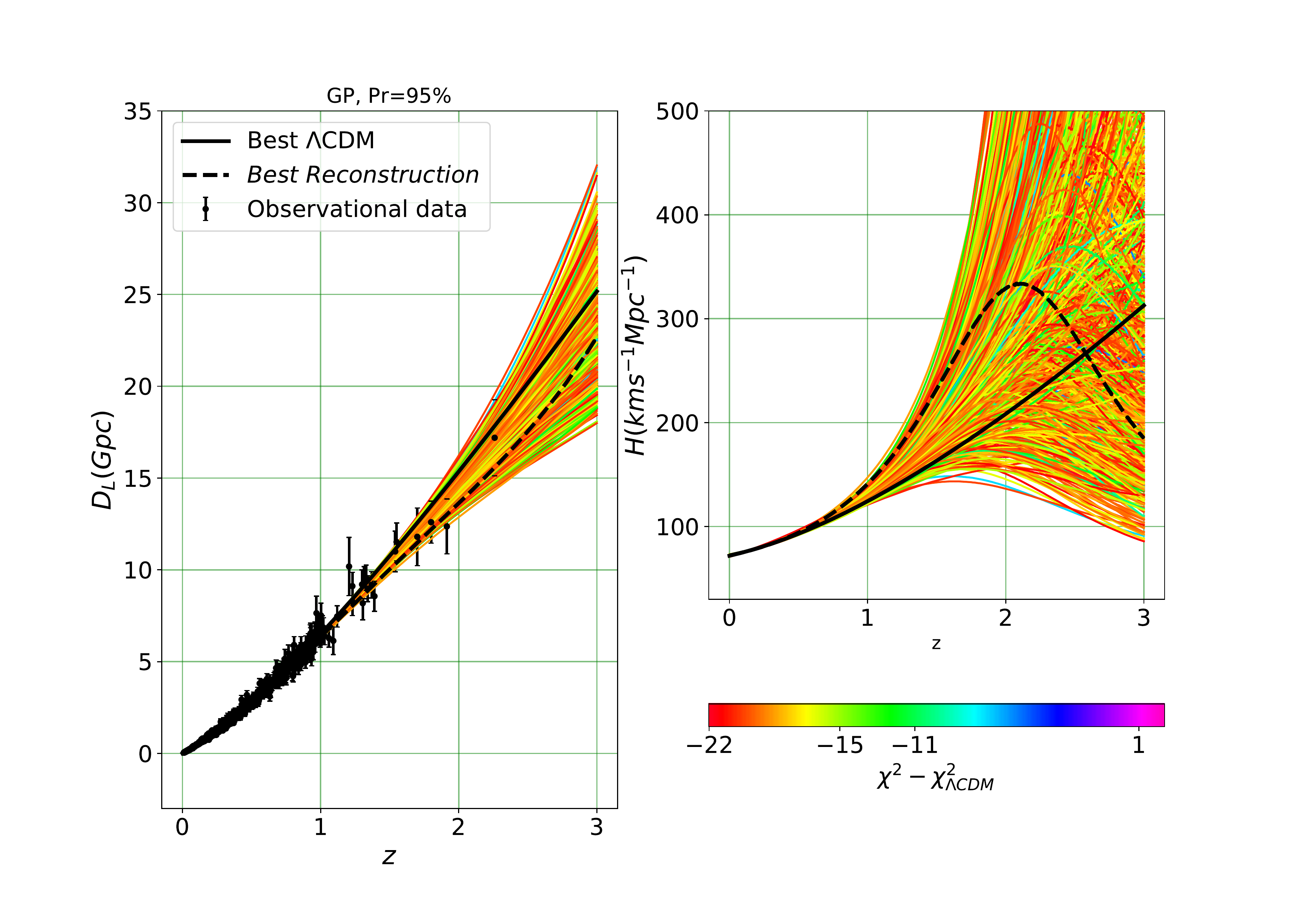}
 	\includegraphics[width=8.2 cm]{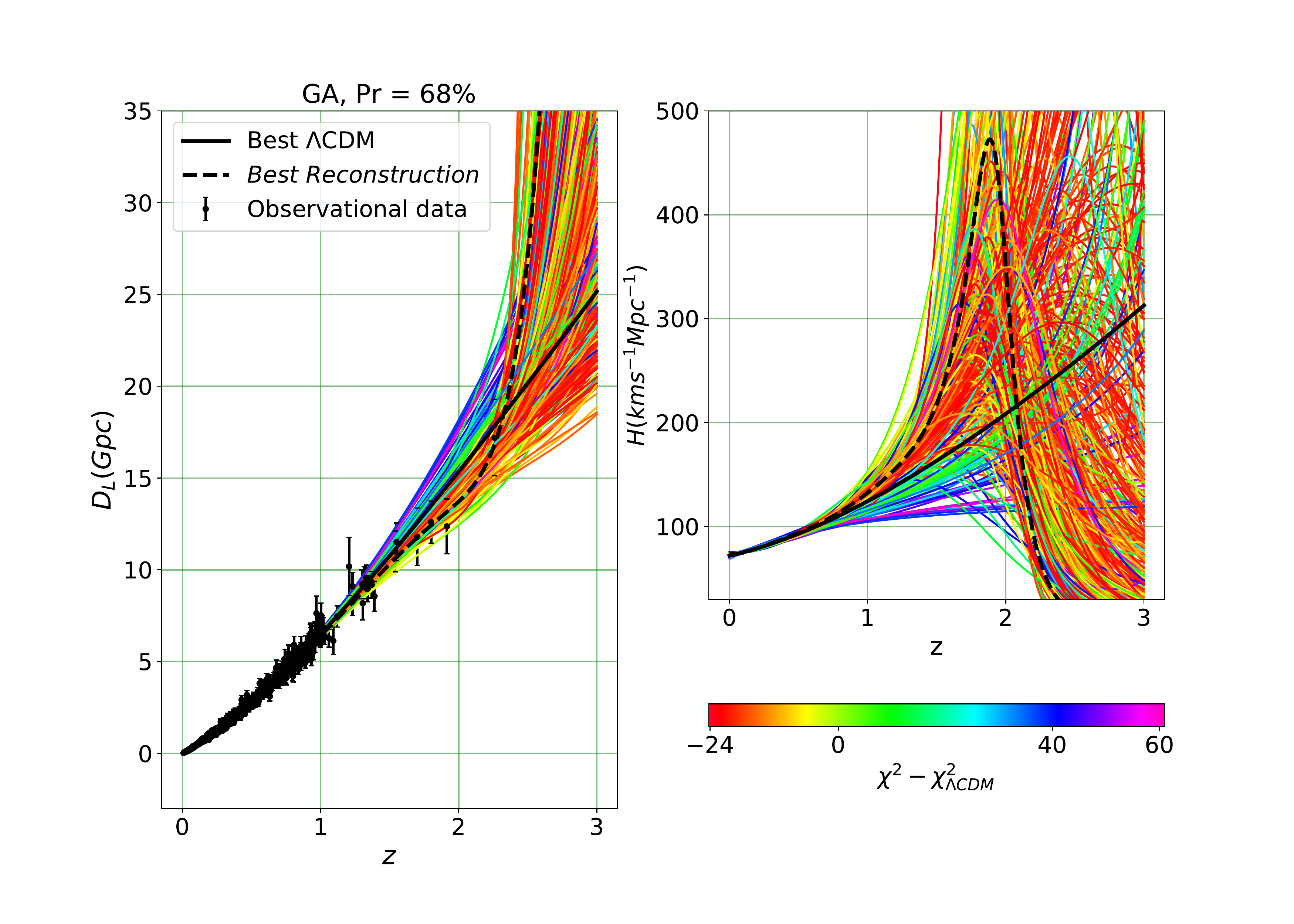}	\includegraphics[width=8.2 cm]{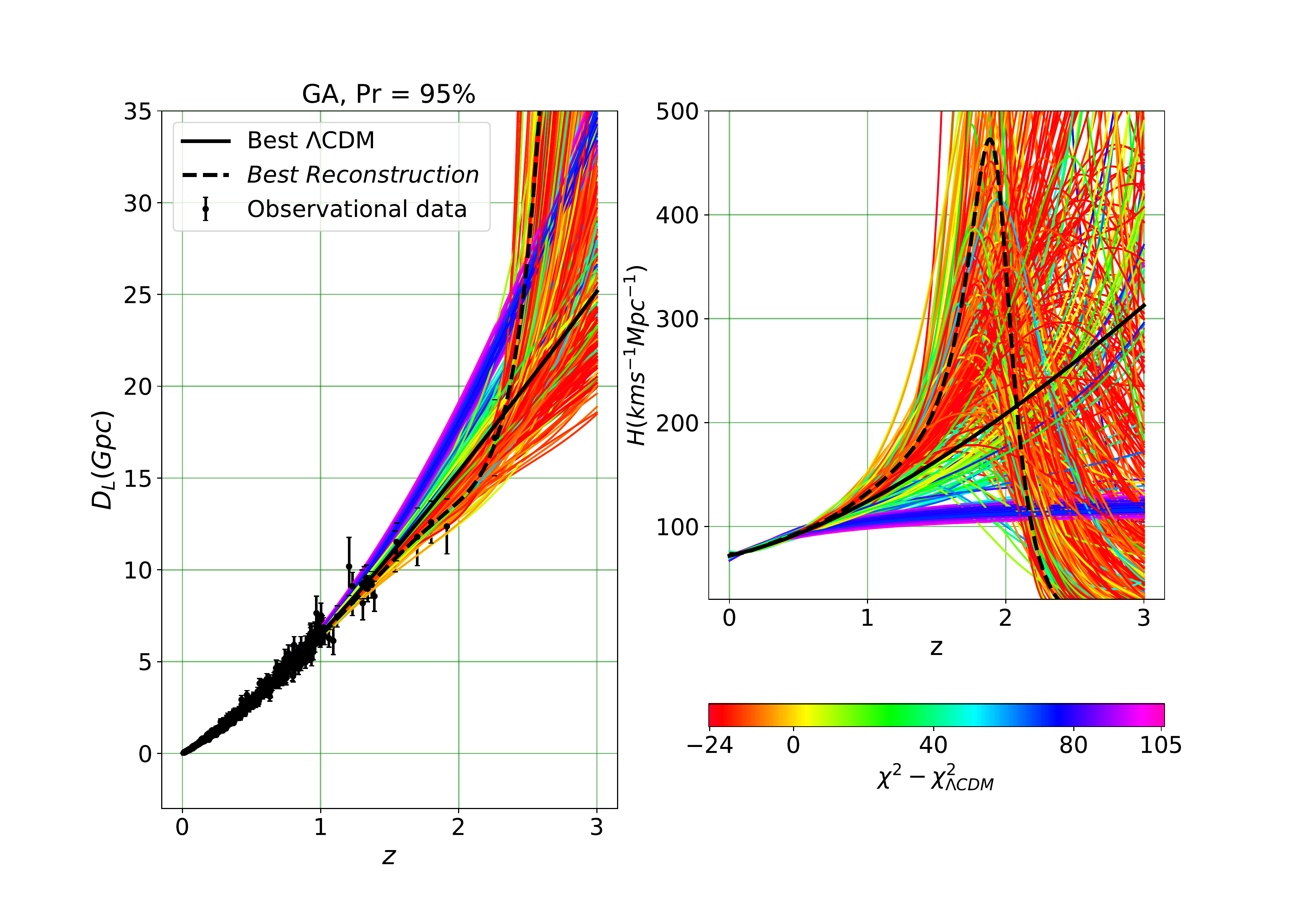}
 	\includegraphics[width=8.2 cm]{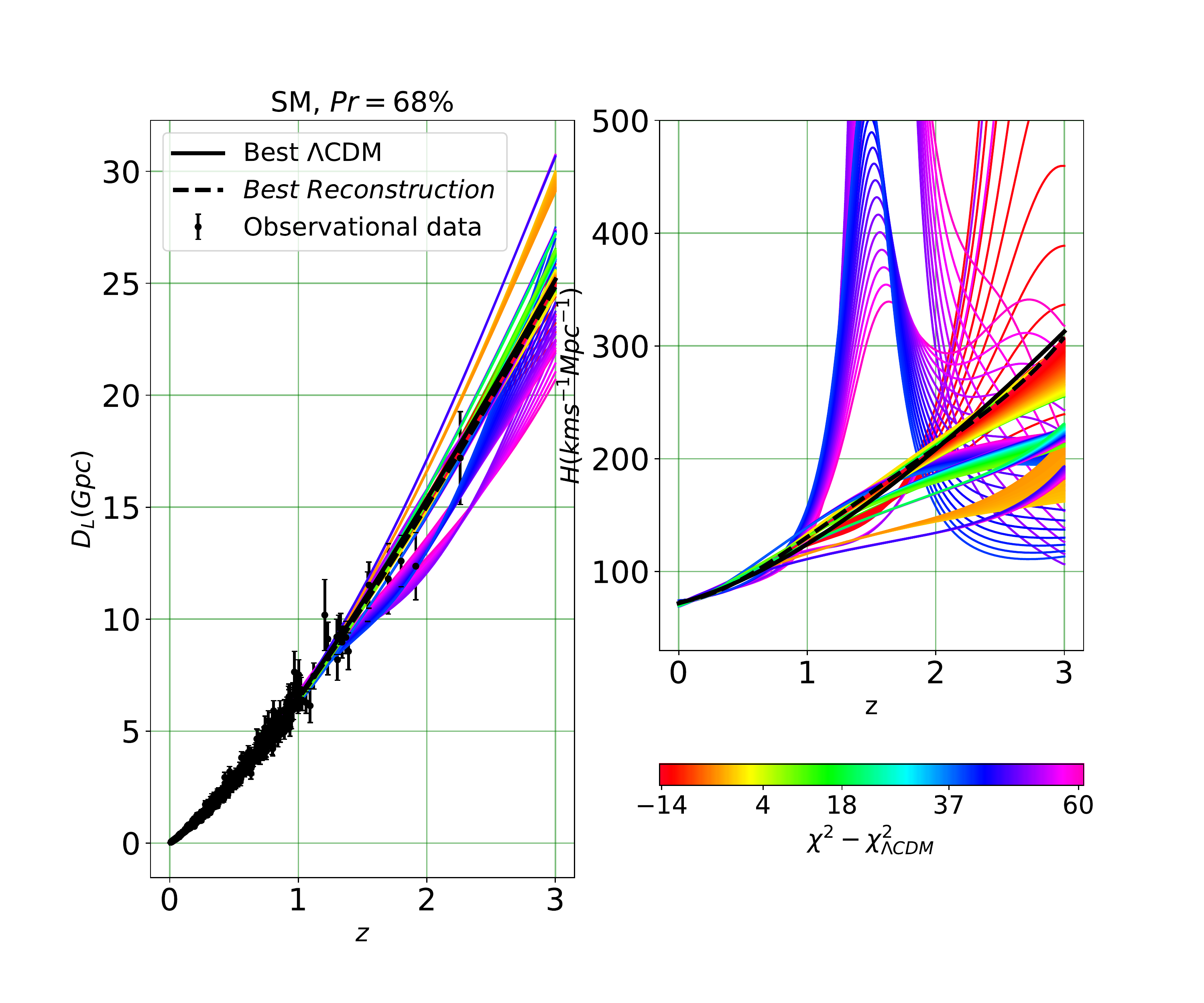}	\includegraphics[width=8.2cm]{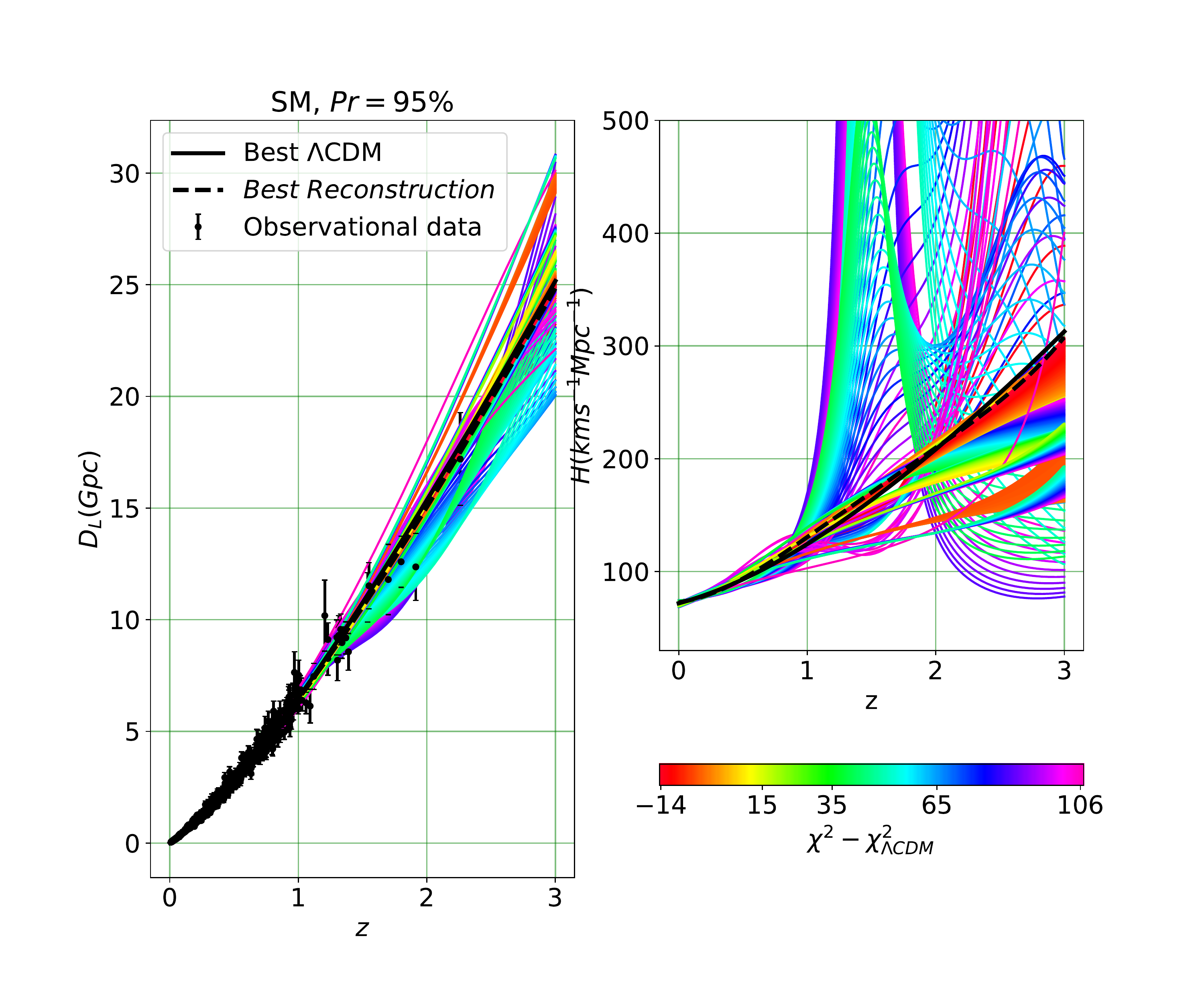}
 	\caption{Reconstructions of the luminosity distance and derived Hubble parameter considering the SNIa data. The upper, middle and lower panel present the results for GP, GA and SM methods respectively. The best $\Lambda$CDM (the best reconstruction) is shown by solid black (dashed black) curve in each panel. The color-bar shows the difference between $\chi^2$ and the best fit $\Lambda$CDM, $\chi^2_{\Lambda CDM}$. In all scenarios the left panel (right panel) presents reconstruction for probability $pr=68\%$ ($pr=95\%$).}
 	\label{fig:all_SN}
 \end{figure*}  
 
As it has been previously stated, methods provide a reconstruction of the luminosity distance in the case of SNIa data, and the Hubble parameter is a derived parameter. In this subsection, we present the results of our analysis using the SNIa data. 
For all techniques, the reconstructions of the luminosity distance and their corresponding Hubble parameter are displayed in Fig. (\ref{fig:all_SN}). Along with all reconstructions, the best fit $\Lambda$CDM and the best reconstructions (the lowest $\chi^2$) have been presented in each panel. The solid black (dashed black) curve in each panel presents the best $\Lambda$CDM (best reconstruction) and the color-bar shows $\Delta\chi^2=\chi^2-\chi^2_{\Lambda CDM}$ quantity. The upper panels show the GP results for $pr=68\%$ and $pr=95\%$ which are almost the same. In fact, all of the reconstructions are within a small range of $\chi2$, so adjusting the PDF probability has little effect on the outcome.

The best reconstructed Hubble has a peak at $z\sim2$ and a smaller value at $z=3$ compare to the best $\Lambda$CDM (for the best reconstruction $\Delta\chi^2=-22$). Notice that almost all of these reconstructed curves have $\chi^2$ values lower than the best $\Lambda$CDM and also the range of $\chi^2$ values is narrower than	in the other two scenarios. 
 
 Considering the GA method, the reconstructed luminosity distance and corresponding Hubble parameter have been presented in the middle panel of Fig.(\ref{fig:all_SN}). The best reconstruction provides a Hubble parameter which has a peak around $z\sim2$ similar to the GP but with even smaller $\chi^2$ value ($\Delta\chi^2=-24$). Furthermore, the results for $pr=68\%$ and $pr=95\%$ are different, with more curves having a larger $\chi^2$ value in the case of $pr=95\%$. These findings suggest that GA provides more flexibility to investigate a data set compare to the GP.

 Finally, the results of SM method have been presented at the lower panel. In contrast to the other methods, the best reconstruction in the SM is very close to the best $\Lambda$CDM. In the SM method, the best reconstruction has a $\Delta\chi^2=-14$, which is relatively larger than other methods. Notice that, since in the case of SNIa the number of data points are larger than the number of Hubble data, a larger $\chi^2$ could be selected based on the chi-squared PDF. For example, in the case of GA and SM the $\Delta\chi^2$ might be large up to 105 while for the Hubble data it is around 25.

\subsection{Results for the Hubble data} 
\begin{figure*}[h]
	\centering
	\includegraphics[width=8.2 cm]{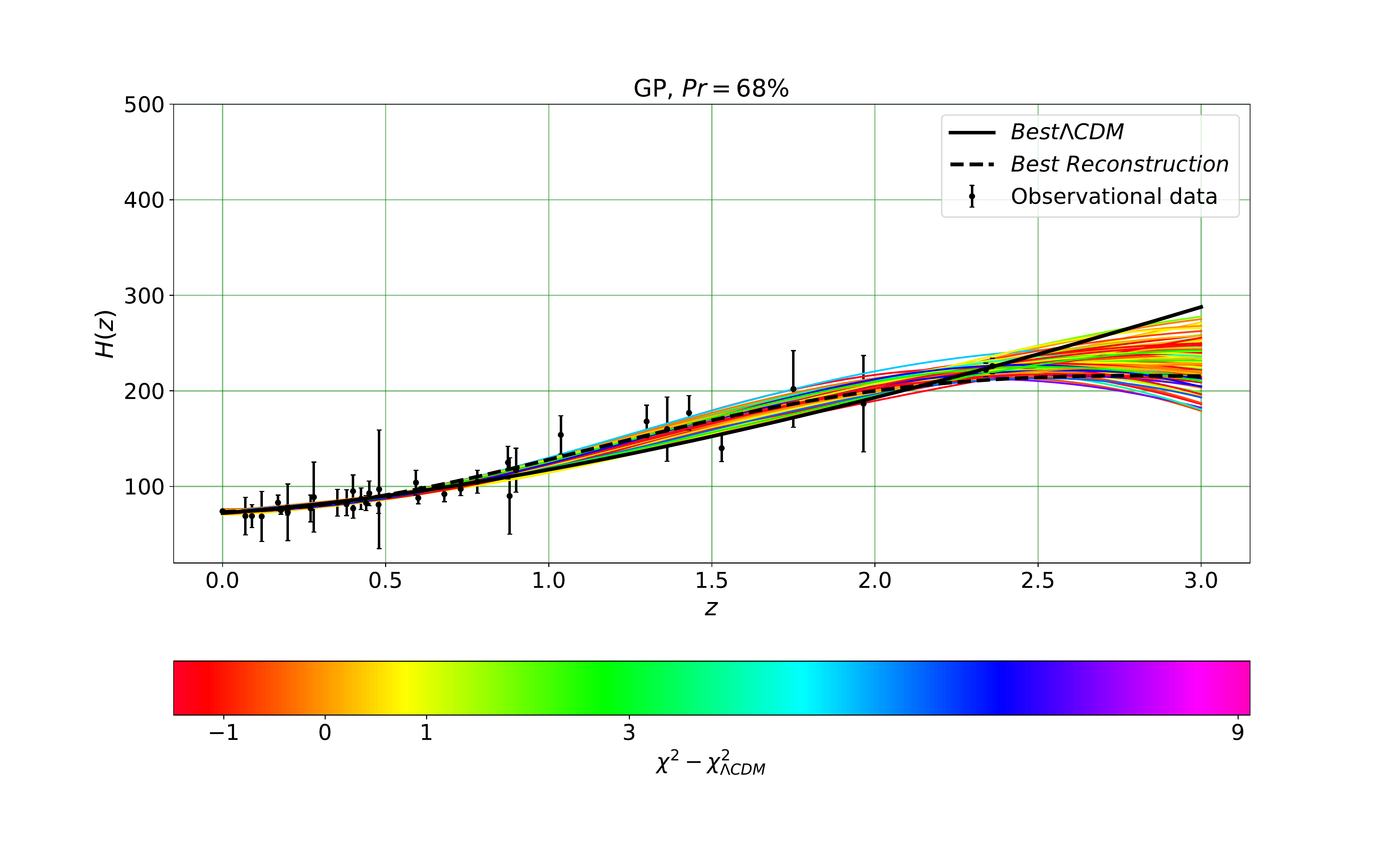}	\includegraphics[width=8.2 cm]{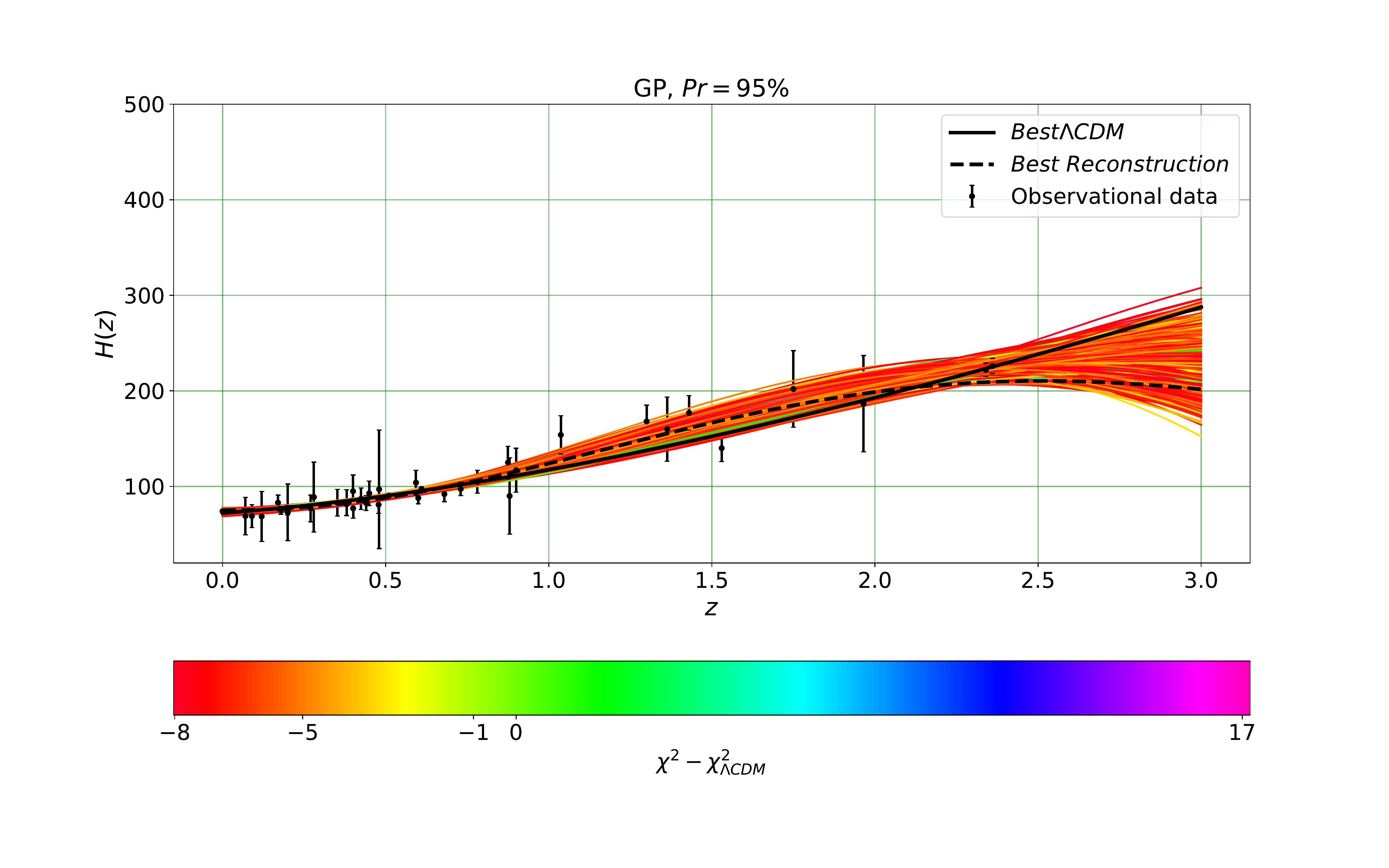}
	\includegraphics[width=8.2 cm]{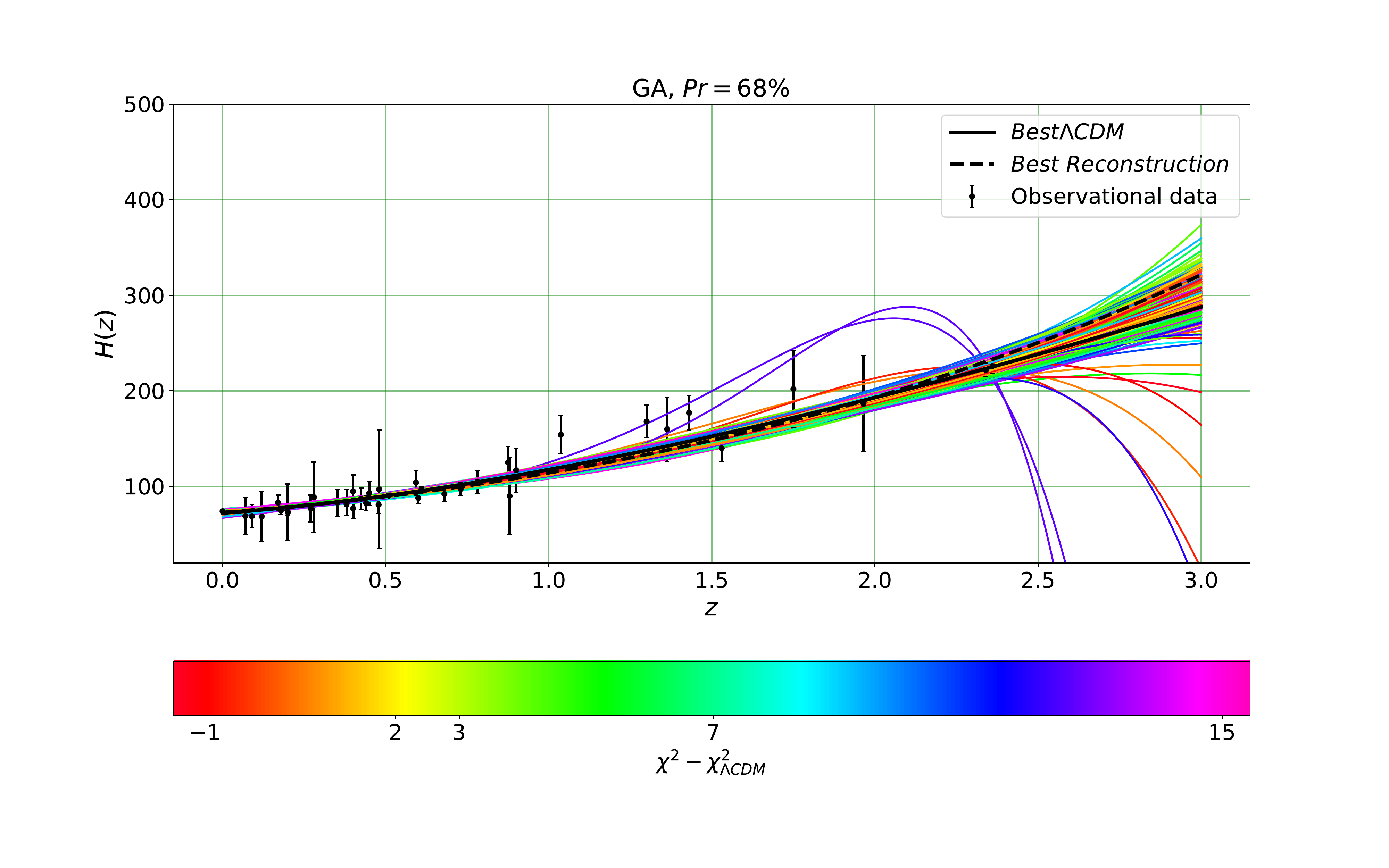}	\includegraphics[width=8.2 cm]{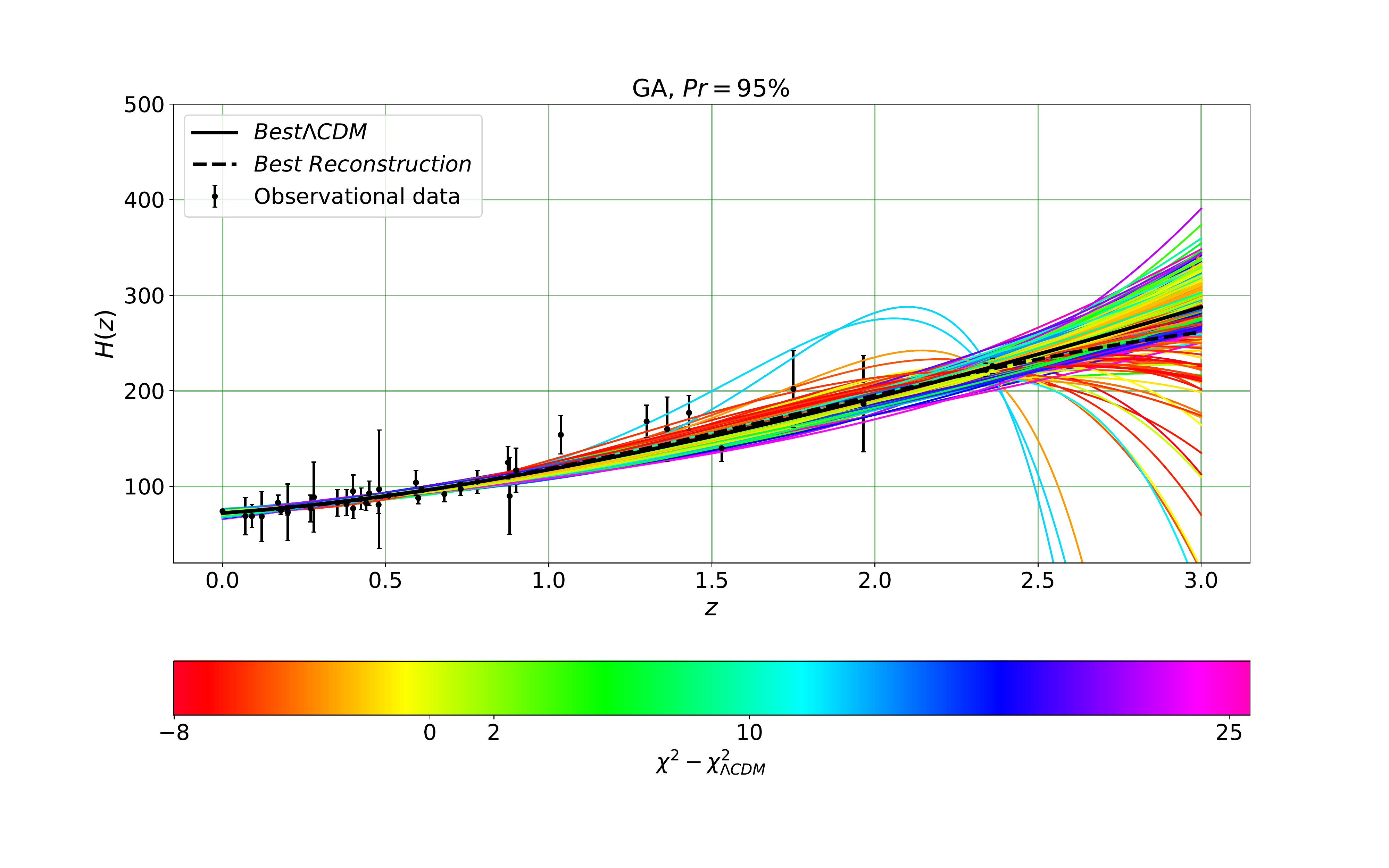}
	\includegraphics[width=8.2 cm]{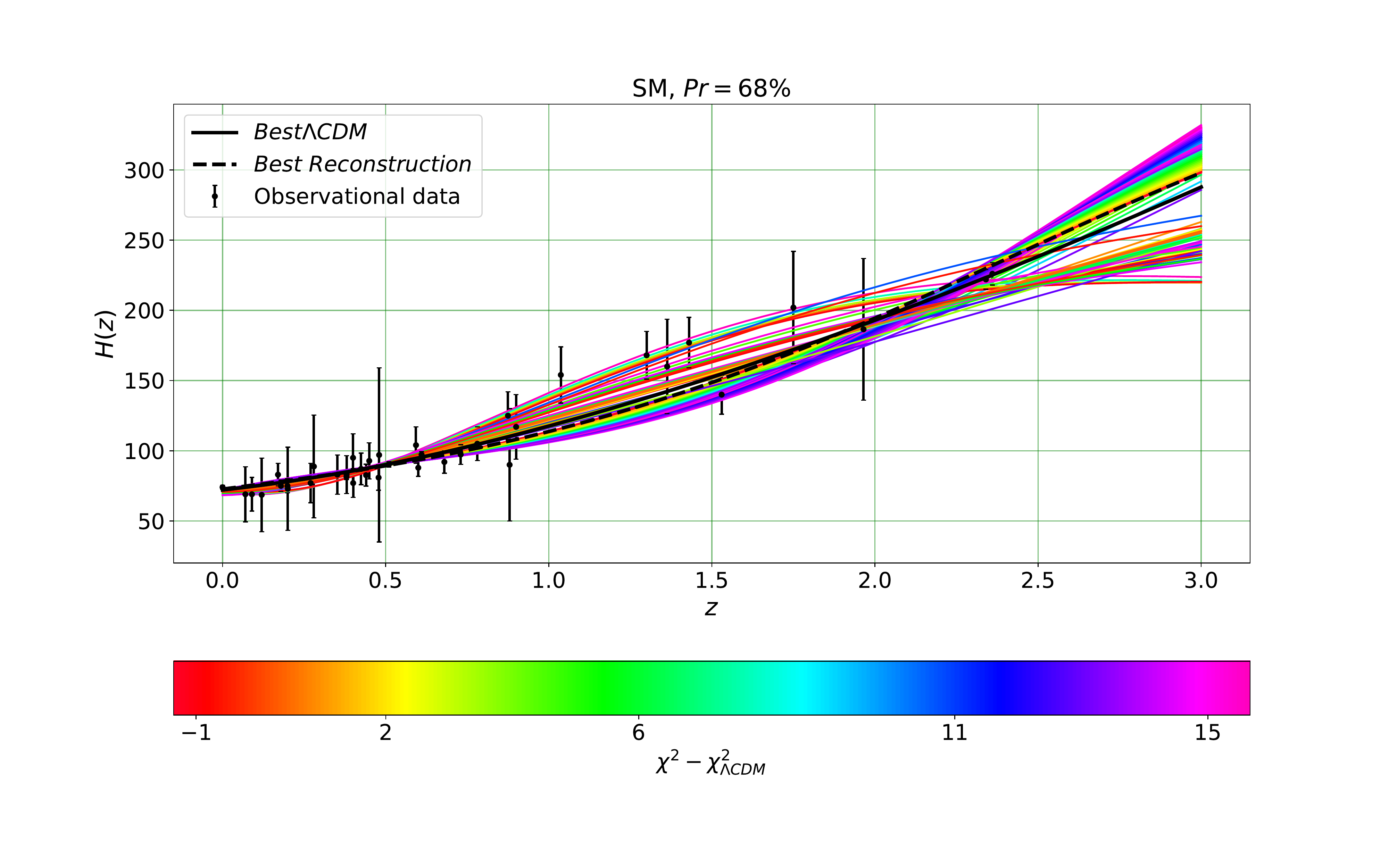}	\includegraphics[width=8.2cm]{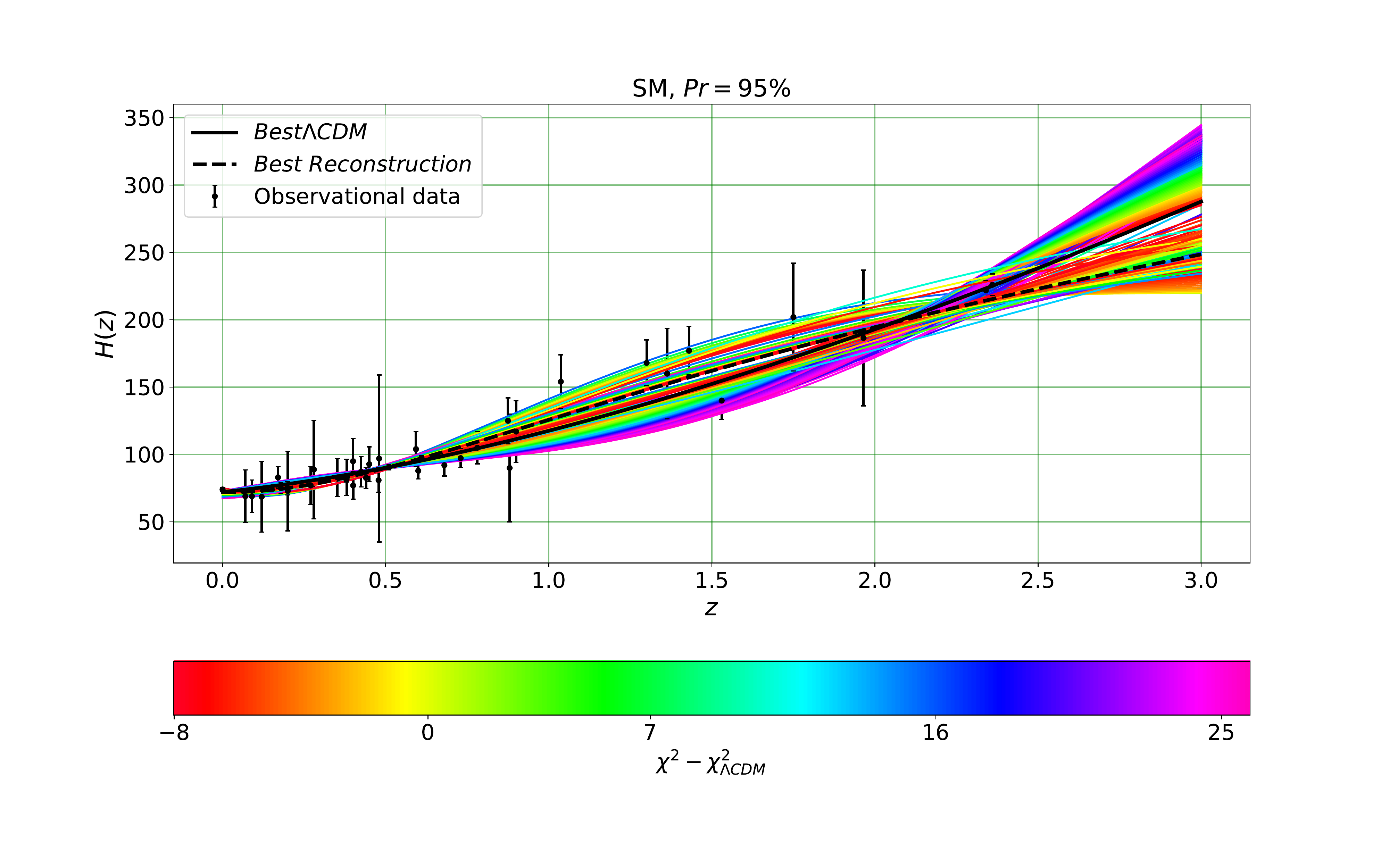}
	\caption{Reconstructions of the Hubble parameter considering the Hubble data. The upper, middle and lower panel present the results for GP, GA and SM methods respectively. The best $\Lambda$CDM (the best reconstruction) is shown by solid black (dashed black) curve in each panel. The color-bar shows the difference between $\chi^2$ and the best fit $\Lambda$CDM, $\chi^2_{\Lambda CDM}$. In all scenarios the left panel (right panel) presents reconstruction for probability $pr=68\%$ ($pr=95\%$).}
	\label{fig:all_H}
\end{figure*}

The reconstructions of the Hubble parameter considering the Hubble data are shown in Fig.(\ref{fig:all_H}) for all the scenarios. The upper panel shows the GP results for $pr=68\%$ and $pr=95\%$ cases. The range of $\chi^2$ in $pr=68\%$ is narrower than $pr=95\%$ as it is expected from the chi-squared PDF. Because of the constraint in the GP (see Eq.(\ref{eq:GP})), a considerable number of reconstructions have a $\chi^2$ around or less than $\chi^2_{\Lambda CDM}$. Moreover, at high redshifts $z\sim 2.5-3$, reconstructions provide a lower $H(z)$ value compare to the $\Lambda$CDM, with the best reconstruction being roughly $25\%$ smaller than the best $\Lambda$CDM at $z=3$. 
Considering the GA method, the results have been shown at the middle panel of Fig.(\ref{fig:all_H}). Similar to the GP, among all reconstructions, only those reconstructions with probability $68\%$ and $95\%$ have been selected and presented in the panels. While the results are similar to the GP up to redshifts $(z<1.5)$, GA provides more scattered curves at $z=3$ and the best reconstruction is closer to the best $\Lambda$CDM than the GP. There are also two curves with $\chi^2-\chi^2_{\Lambda CDM}\sim10$ that have a minor peak around $z=2$. These results suggest that GA may be more useful and adaptable than GP when it comes to discovering a new feature in a data set. 

Finally, the results of SM method are displayed in the lower panel. The $\chi^2$ range is similar to that of GA, although there are more curves with large $\chi^2$. Moreover, reconstructions with a high $\chi^2$ had a lower $H(z)$ (purple curves) at $z\sim 1.5$ and a higher value at $z=3$. The best reconstruction is nearly identical to the best $\Lambda$CDM up to redshift $z\sim2$ but gives a slightly larger (smaller) value in the case $pr=68\%$ ($pr=95\%$) at redshift $z=3$.

Our findings suggest that using the chi-squared PDF to select reconstructions provides a better insight in analyzing a data set and in this scenario some reconstructions may reveal a new feature. In contrary, selecting reconstructions with $\chi^2$ smaller than a threshold value may not offer a comprehensive sample. It is important since in the majority of cases our main objective is the estimation of a quantity (like the $H_0$) from the sample.       

\subsection{Estimation of $H_0$}

\begin{figure*}[h]
	\centering
	\includegraphics[width=8 cm]{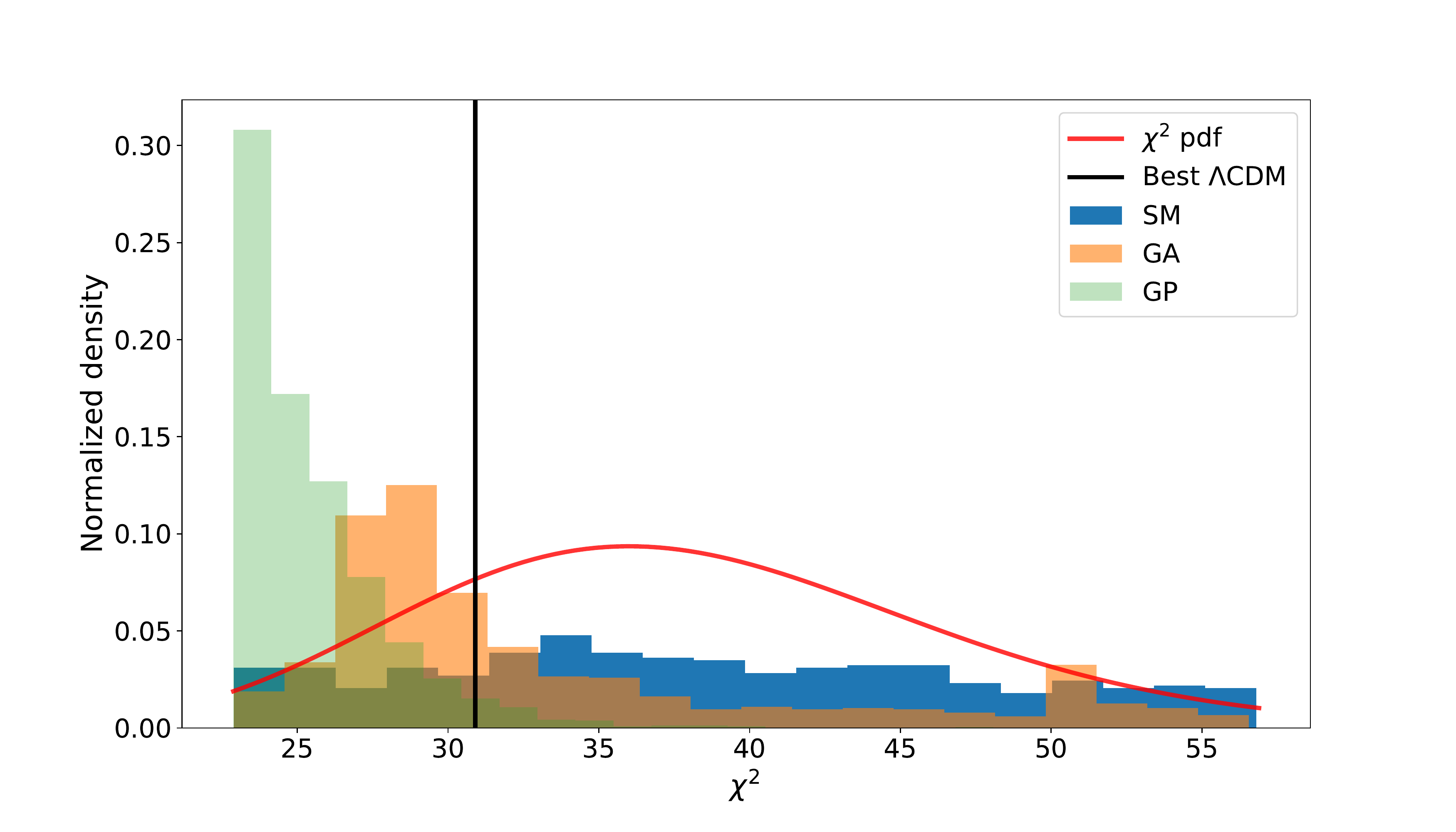}	\includegraphics[width=8 cm]{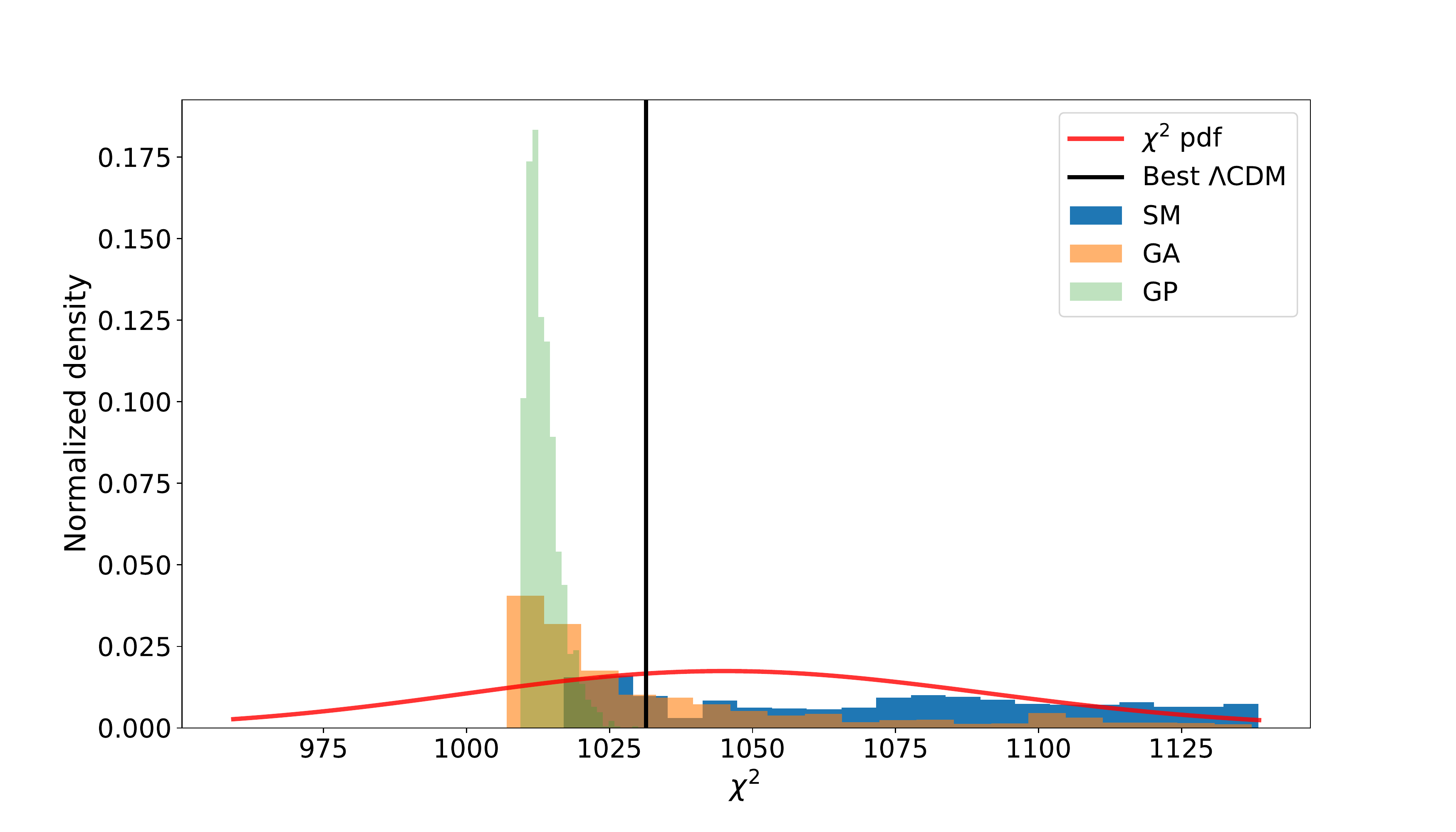}
	\caption{Distribution of the reconstructions $\chi^2$ in each methods for $pr=95\%$. The solid red line indicates the $\chi^2$ PDF corresponding to the number of degree of freedom for each data set. Left panel (Right panel) shows the results for the Hubble data( the SNIa data). }
	\label{fig:chi2}
\end{figure*}

For both Hubble and SNIa data, the distribution of $\chi^2$ for all reconstructions, as well as the $\chi^2$ PDF, is presented in Fig.(\ref{fig:chi2}). In these diagrams, the probability has been set to $pr=95\%$, the red solid line indicates the $\chi^2$ PDF and the vertical black line shows the location of the best $\Lambda$CDM. For both data sets, the SM produces a nearly  uniform distribution but GA and GP provide more reconstructions with smaller $\chi^2$. Notice that, the results of SM (GA) are dependent on the number of iterations (generations) and raising this number results in more reconstructions with less $\chi^2$ (albeit $\chi^2$ does not reduce significantly after some steps).The distribution of $\chi^2$ in GP, on the other hand, covers a narrow region and is unaffected by sample size. Moreover, whereas the Hubble data reconstruction cover the whole range of $\chi^2_{min}<\chi^2<\chi^2_{max}$ for $pr=95\%$, the SNIa reconstructions are unable to reach the minimum value $\chi^2_{min}$, and all GP reconstructions have a $\chi^2$ smaller than the best $\Lambda$CDM. 

As we mentioned above, the reconstructions may be used to estimate the value of $H_0$ as well as its uncertainty in a model independent manner. In fact, estimating the central value is an easy task but estimating the uncertainty is more difficult. While the central value can be easily estimated from the mean or median, there are different approaches for estimation of the uncertainty. Considering the SM method, in \cite{Shafieloo_2007,Shafieloo_2010,Shafieloo_2012_2,10.1093/mnras/sty398,Shafieloo_2018} authors used the maximum and minimum values at each redshift to find an interval of the uncertainty. In the GA scenario, on the other hand, the path integral approach has been used to estimate the uncertainty \cite{Nesseris_2012,Nesseris_2013}. Note that, in this case the method gives only the uncertainty of the best reconstruction not the overall uncertainty. In a GP approach, however, estimate of these quantities is simple, and the mean (standard deviation) at each redshift gives the central value (uncertainty). In order to estimate the value of $H_0$ as well as its uncertainty, we use a similar process as GP.

\begin{figure*}[h]
	\centering
\includegraphics[width=8 cm]{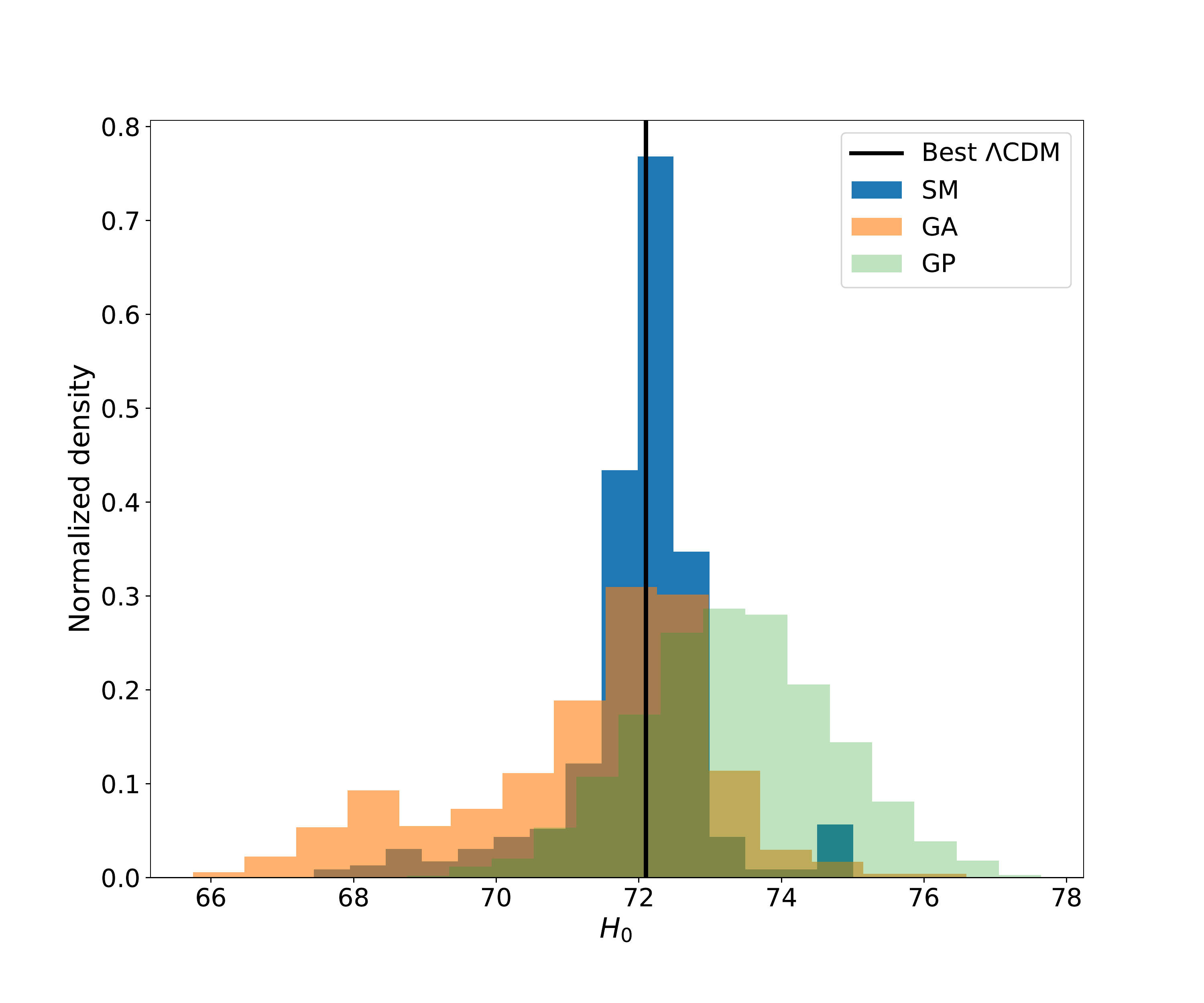}	\includegraphics[width=8 cm]{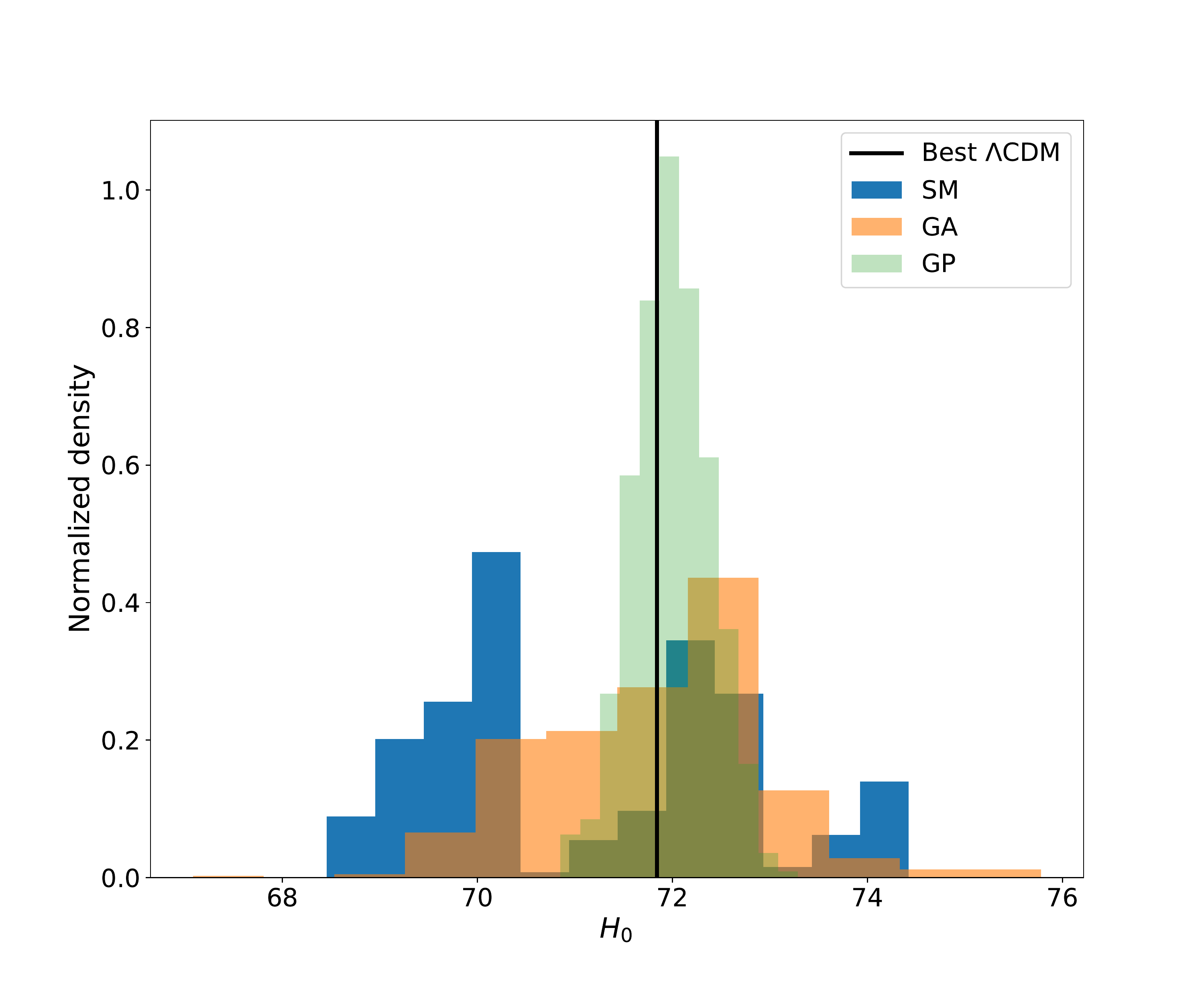}
	\caption{The distribution of estimated $H_0$ considering $pr=95\%$ in different methods. The left panel (right panel) shows the results for the Hubble parameter (the SNIa data).}
	\label{fig:H0}
\end{figure*}

 The distributions of $H_0$  for both data sets are illustrated in Fig.(\ref{fig:H0}). The solid vertical line shows the location of the best $\Lambda$CDM and distributions of $H_0$ from different methods are presented by different colors. For the Hubble data, the SM method provides a peak near to the $\Lambda$CDM, whereas results from GA and GP are scattered over a large area. In contrast to the other two approaches, the peak of the distribution in GP gives a substantially greater value $H_0\sim 74$. This is a direct consequence of the SHOES \cite{Riess:2019cxk} data point which shift $H_0$ towards a larger value. Since the GP has been widely utilized in the literature, it is important to remember that one data point (especially one with a small error) can significantly alter the results. Unlike the GP, GA is unaffected by this data point and provides a relatively smaller $H_0$. On the other hand, for the SNIa data, we see a narrow distribution around the value of $\Lambda$CDM in GP and a relatively wider distribution for both SM and GA.

\begin{figure*}[h]
	\centering
	\includegraphics[width=9 cm]{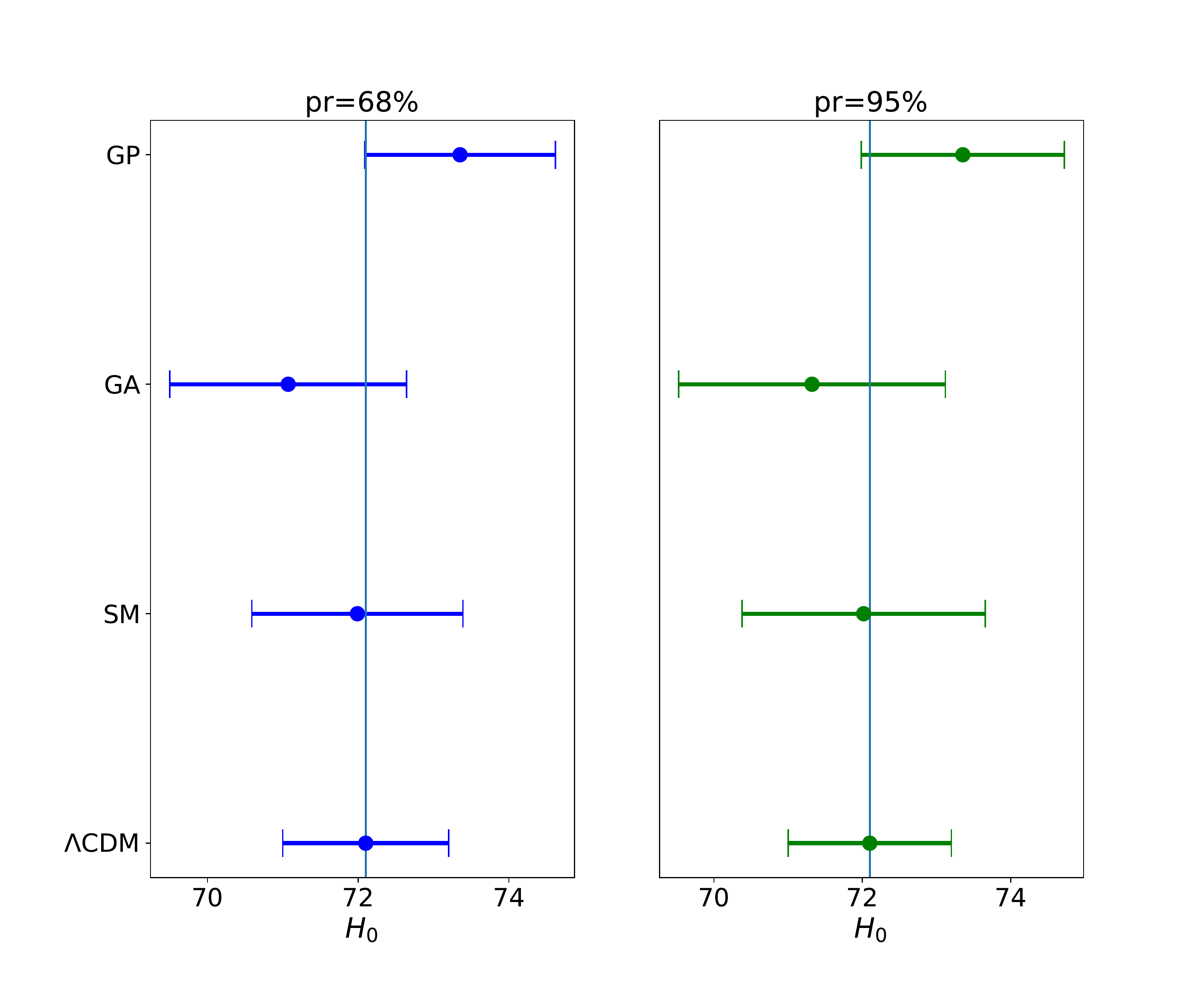}\includegraphics[width=9 cm]{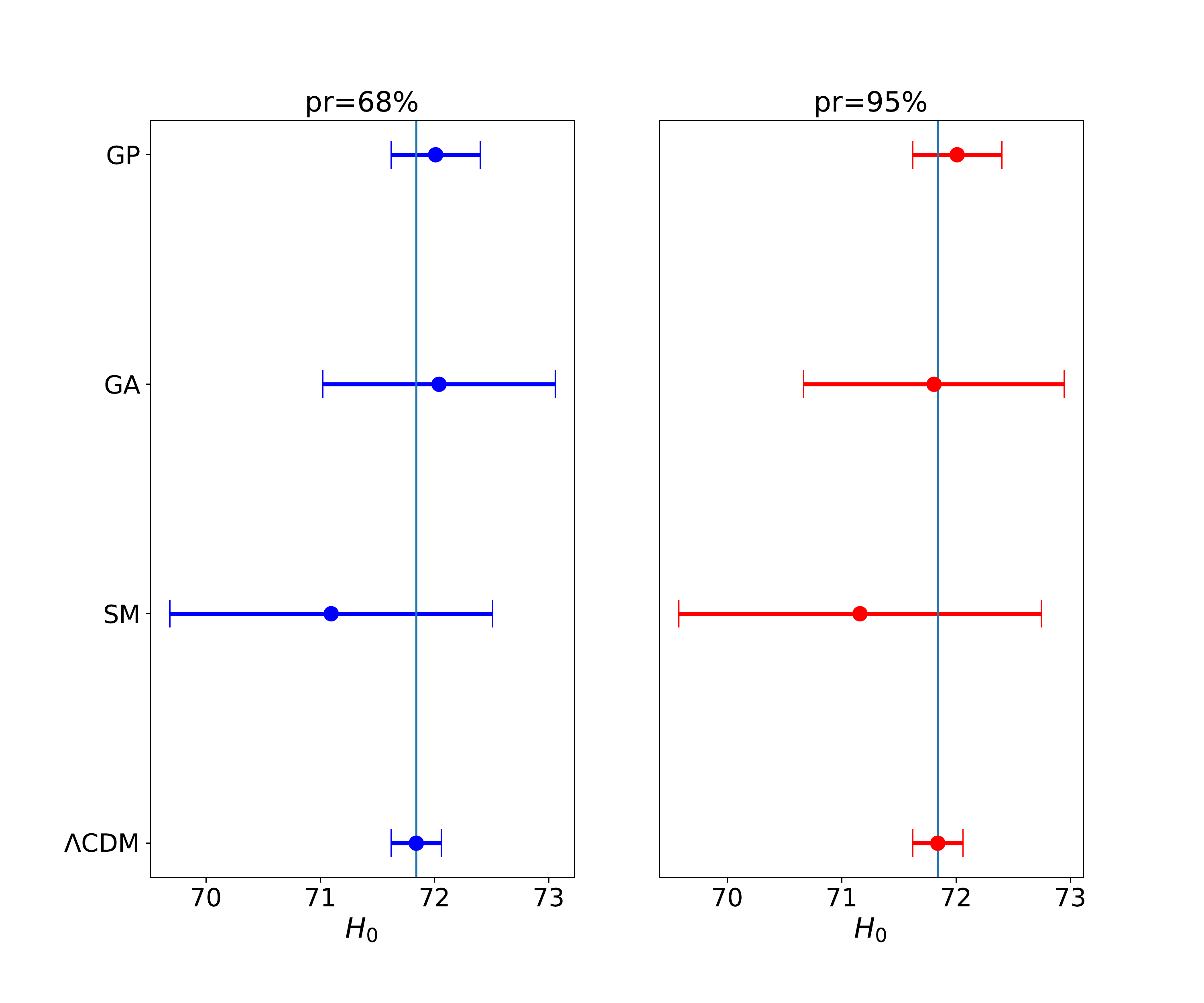}
	\caption{Estimation of $H_0$ and its $1\sigma$ uncertainty in GP, GA and SM for $pr=68\%$ and $pr=95\%$. The left panel (right panel) shows the results for the Hubble data (the SNIa data).}
	\label{fig:H0_val}
\end{figure*}

In contrast to Hubble data, the results of GP in the case of SNIa show a tight peak around the best $\Lambda$CDM, indicating that the central value is close to the $\Lambda$CDM and its uncertainty is smaller than other methods. in this case, the SM method yields a relatively wider distribution and we have $H_0$ in range of ( 68.5-74.5). Since the results might depend on the probability in our selection criterion, we perform our analysis with both $pr=68\%$ and $pr=95\%$. The results are presented schematically in Fig.(\ref{fig:H0_val}) and quantitatively in Tab.(\ref{tab:results}).
 The main points regarding the $H_0$ estimation are as follows:
\begin{itemize}
	\item The results of all methods are consistent with  $\Lambda$CDM at 1$\sigma$ level.
	\item The results for $pr=68\%$ and $pr=95\%$ are consistent with each other but uncertainties are around $10-20\%$ larger for $pr=95\%$ in the GA and SM.
	\item The uncertainties in the GP for the SNIa data are the same for both $pr=68\%$ and $pr=95\%$. This is mainly due to the fact that reconstructions in the GP are concentrated in a small area of the chi-squared PDF, which does not change with probability. On the other hand, $pr=68\%$ offers a $6\% $ lower uncertainty for Hubble data than $pr=95\% $. 
	\item The SHOES data point shifts the $H_0$ towards a larger value in the GP but other two methods are not sensitive to this data point.
	\item The GP has the least uncertainty among all the methods and uncertainties in the GP are only slightly larger than the $\Lambda$CDM. 
	
\end{itemize}
  

\begin{table}
	\centering
	\begin{tabular}{|l | c | c |}
		\hline 
	Method/data	&  Hubble data &  SNIa        \\
		\hline \hline 
		SM (Pr=68$\%$)  & $71.99\pm1.43$ & $71.09\pm1.41$\\
		\hline 
		SM (Pr=95$\%$)  & $72.02\pm1.64$& $71.16\pm1.59$\\
		\hline \hline  
		GA(Pr=68$\%$) & $71.07\pm1.57$&$72.04\pm1.02$ \\
		\hline 
		GA(Pr=95$\%$) & $71.32\pm1.80$ &$71.80\pm1.14$\\
		\hline \hline 
		GP(Pr=68$\%$)&$73.33\pm1.27$ & $72.02\pm0.38$ \\
		\hline
		GP(Pr=95$\%$) &$73.35\pm1.35$& $72.02\pm0.38$ \\
		\hline
		$\Lambda$CDM&$72.1\pm1.1$& $71.85\pm0.22$ \\
		\hline \hline
			\end{tabular}
	\caption{Estimation of $H_0$ (km/s/Mpc) and its $1\sigma$ uncertainty in GP, GA and SM for $pr=68\%$ and $pr=95\%$. The left column (right column) shows the results for the Hubble data (the SNIa data).  }\label{tab:results}
\end{table}

\section{Conclusion}\label{conclude}
 In this work, we consider three well-known NP methods, all reliant upon a sample of reconstructions, namely GP, GA and SM methods and introduce a novel approach to select a consistent reconstruction. We compare the results of employing the SNIa (the Pantheon sample) and a recent collection of Hubble parameters to the NP techniques and check the consistency of our selection criterion. Unlike previous studies, we use the probability of each reconstruction based on the chi-squared PDF to select a consistent reconstruction. Given the NDF of a data set, it is straightforward to find the probability of each reconstruction (according to its $\chi^2$ value) and setting a probability threshold, such as $pr=68\% $ or $pr=95\% $, one can separate reconstructions with these probabilities according to the chi-squared PDF.
However, a different strategy has utilized in \cite{Shafieloo_2007,Shafieloo_2010,Shafieloo_2012_2,10.1093/mnras/sty398,Shafieloo_2018} to select a reconstruction. In these works authors have considered a reference $\chi^2_{\rm ref}$ as a threshold and selected all reconstructions with a $\chi^2$ smaller than the threshold. 

 Considering the Hubble data, we reconstruct the Hubble parameter directly and compare the results for the cases of $pr=68\%$ and $pr=95\%$ in all three methods. On the other hand, the Hubble parameter is computed from the reconstructed luminosity distance for the SNIa data. According to our results, the GA is more flexible in finding a new feature compare to the other methods. Moreover, for both data sets, GP provides a smaller range of $\chi^2$ with the majority of them having a $\chi^2$ smaller than the concordance $\Lambda$CDM. In addition, our analysis indicates that GP is more efficient than other two methods in terms of computational time, whereas SM is the slowest.  

The reconstructions could also be used to estimate the value of $H_0$, an essential quantity in cosmology. It is worth noting that estimating the $H_0$ requires a reliable sample of reconstructions. For all reconstructions in the sample, we obtain the $H_0$ and investigate its distribution for each data set considering all methods. Notice that the estimating uncertainty in an NP technique is a difficult task.

 Specifically,  in \cite{Shafieloo_2007,Shafieloo_2010,Shafieloo_2012_2,10.1093/mnras/sty398,Shafieloo_2018}, authors have used the maximum and minimum values at each redshift to determine an uncertainty interval. The authors of \cite{Nesseris_2012,Nesseris_2013,Arjona:2019fwb,arjona2020machine}, on the other hand, have employed a path integral-based method to estimate the uncertainty in the GA.
 In this paper, we adopt the same procedure as GP and estimate the central value using the sample main at each redshift and the uncertainty using the standard deviation. Our results indicate that GP has the lowest uncertainties which are slightly higher than those found in the concordance $\Lambda$CDM. On the other hand, while all the estimated central values are consistent with each other, in SM and GA, the estimation of uncertainty for  $pr=95\%$ is roughly $10-20\%$ more than the results for $pr=68\%$. Furthermore, changing the chi-squared probability has no significant effect on the estimated uncertainty in GP for both data sets. This is primarily owing to the fact that the GP gives a limited range of $\chi^2$ compare to the other two methods. Based on our findings, it may be preferable to use the GA or SM in conjunction with the GP when studying a data set.
  
Estimation of $H(z)$ in a model independent manner is not new. In particular, authors of \cite{Lemos:2018smw} have combined the Pantheon and BAO measurements and utilized a parametric form of $H(z)$ to find the value of $H_0$. They have assumed some priors on the sound horizon at the drag epoch and obtained two $H_0$ estimates that were significantly closer to the Planck $\Lambda$CDM estimate than the SHOES estimation. Our estimate is around $1.5-3\sigma$ larger than theirs, owing to the fact that our results are independent of the sound horizon.
On the other hand, our results are very similar to those presented in \cite{Gomez-Valent:2018hwc}, which used the Hubble data alone as well as a combination of Hubble data and the SNIa to estimate $H_0$ from a GP. In addition, authors of \cite{Arjona:2019fwb}, employed GA to estimate the $H_0$ using similar data set. They found a considerably larger uncertainty $\sim 12$ km/s/Mpc using the path integral approach, than what we obtained $\sim 2$ km/s/Mpc.     

 Finally, our results indicate that the SHOES data point has a significant impact on estimation of $H_0$ in GP, whereas other two methods are less affected by this data point. In fact, when GP is used instead of the other two approaches, the data point shifts the peak of the $H_0$ distribution towards a bigger value. This finding also suggests that considering GA or SM along with the GP may be beneficial.
 

 \bibliographystyle{apsrev4-1}
  \bibliography{ref}

\end{document}